\documentclass[preprint,12pt,authoryear]{elsarticle}
\usepackage[top=2cm,bottom=2.5cm,left=2.5cm,right=2.5cm]{geometry}

\usepackage{amssymb}
\usepackage{amsmath}
\usepackage{mathdots}

\graphicspath{{figures/}}
\usepackage{subfig}
\usepackage{stackengine}
\usepackage{booktabs}	
\usepackage{tikz}

\usepackage{color}
\usepackage[normalem]{ulem}

\usepackage{siunitx}
\let\DeclareUSUnit\DeclareSIUnit

\DeclareUSUnit\pound{lb}
\DeclareUSUnit\poundforce{lb\textsubscript{f}}
\DeclareUSUnit\foot{ft}
\DeclareUSUnit\gallon{gal}

\newcommand\Rey{\mbox{\textit{Re}}}  

\usepackage[nolist,nohyperlinks]{acronym}
\begin{acronym}
	\acro{PSM}{partially saturated method}
	\acro{LBM}{lattice Boltzmann method}
	\acro{DEM}{discrete element method}
	\acro{TRT}{two-relaxation-time}
	\acro{FEM}{finite element method}
	\acro{DFM}{diffusive flux model}
	\acro{DNS}{direct numerical simulation}
	\acro{PSD}{particle size distribution}
	\acrodefplural{PSD}[PSDs]{particle size distributions}
	\acro{HF}{hydraulic fracturing}
	\acro{CSG}{coal seam gas}
	\acro{SVF}{solid volume fraction}
	\acrodefplural{SVF}[SVFs]{solid volume fractions}
\end{acronym}

\usepackage{hyperref}

\journal{Powder Technology}

\begin{document}
	
	\begin{frontmatter}
		
		
		
		\title{Proppant transport at the intersection of coal cleats and hydraulic fractures}
		
		\author[a,b]{Nathan J. Di Vaira}
		\author[a,c]{Łukasz Łaniewski-Wołłk}
		\author[b]{Raymond L. Johnson Jr.}
		\author[a]{Saiied M. Aminossadati}
		\author[a,b]{Christopher R. Leonardi}
		
		\affiliation[a]{organization={School of Mechanical and Mining Engineering, The University of Queensland},
			city={Brisbane},
			postcode={4072},
			country={Australia}}
			
		\affiliation[b]{organization={Gas and Energy Transition Research Centre, The University of Queensland},
			city={Brisbane},
			postcode={4072},
			country={Australia}}
		
		\affiliation[c]{organization={Institute of Aeronautics and Applied Mechanics, Warsaw University of Technology},
			city={Warsaw},
			postcode={00-665},
			country={Poland}}

		\begin{abstract}
			
			This work is the first computational study of proppant leak-off through coal cleats that accounts for proppant retention in cleats, occlusion formation at cleat entrances, the resulting control of fluid leak-off, and the influence of realistic cleat roughness on these factors. The fluid-particle suspensions are directly simulated with a coupled lattice Boltzmann method-discrete element method which explicitly models all pertinent physics, including shear-thinning fluid rheology. Firstly, using a simplified computational geometry, it is demonstrated that leak-off and mounding are both minimised when proppant invades and is retained in the cleat. This occurs most effectively with wide proppant size distributions, such as 100/635 mesh. When the proppant is larger than the cleat aperture, however, occlusions form at the cleat entrance which can lead to significant mounding, which is observed for 100 mesh and 40/70 mesh. These findings are commensurate with an existing benchmark experiment. Furthermore, mounding is significantly reduced for shear-thinning fluids compared to Newtonian fluids. Simulations are then conducted with rough cleats in a computational geometry that simulates a small fraction of the total hydraulic fracture. Leak-off is smallest for high cleat roughness and when cleats are narrower than a critical width, while proppant retention is largest at the same critical width but only above a certain roughness. Mounding is primarily dependent on the width, as opposed to the roughness. These results are presented as high-fidelity maps which can be directly incorporated into hydraulic fracturing simulations for improved predictions of fluid leak-off and propped reservoir volumes, and which can be tailored for different treatment and reservoir conditions.
			
		\end{abstract}

		
		
		
		
	\end{frontmatter}
	
	
	\section{Introduction}
	\label{Section: Introduction}
	
	Micron-scale proppant, such as 100 mesh and even smaller silica flour (635mesh) and industrially-synthesised ceramics, have recently seen an increase in use for the stimulation of \ac{CSG} reservoirs, under the assumption that they penetrate the pre-existing cleat network more effectively than conventionally-sized proppant. In general, however, the success seen in some treatments is not universally repeatable. This variability is primarily due to the limited observations which can be made in the field, combined with the complex physical phenomena which are present in hydrodynamic flows of particle-laden suspensions. Recently, for example, novel migration behaviour associated with the poor size control of micro-proppant has been demonstrated~\citep{DiVaira2022}.
	
	For proppant to effectively prop the cleat network, it must both invade the cleats from the hydraulic fracture and be retained in the cleats. At the same time, it is desirable for proppant to travel far into the cleat network before being retained, for maximum cleat stimulation. Proppant also has the function of controlling fluid leak-off into the cleats, via either retention in the cleats adjacent to the hydraulic fracture or, if the proppant is too large relative to the cleat aperture, the formation of occlusions at the cleat entrance. These occlusions, however, can result in significant mounding at the cleat entrance, which can potentially block most of the hydraulic fracture~\citep{Penny1991,Stimlab1995}. On the other hand, if proppant is too small it will flow straight through the cleats, meaning that leak-off is not controlled and the cleat is not propped~\citep{Stimlab1995}. The shape and size of the cleat apertures, which can be three orders of magnitude smaller than the hydraulic fracture, is a critical determinant of this size exclusion, and therefore of leak-off. 
	
	In the period from 1990 through 1995, STIM-LAB Inc. conducted a wide-ranging study into various aspects of \ac{HF} treatments of \ac{CSG} reservoirs. The final report~\citep{Stimlab1995} contains the outcomes and recommendations of a number of novel experiments, ranging from the stimulation benefits of various fluid types and chemical additives, to the placement of proppant, to the impact of these treatments on the resulting reservoir permeability. In particular, the report represents the state of the art for modelling proppant transport through hydraulic fractures with intersecting cleats. It is the only work, of which the authors are aware, that clearly demonstrates how proppant can be used to control fluid leak-off in coals, as well as the mounds of proppant which can consequently develop in the hydraulic fracture. The present paper therefore takes the STIM-LAB report as a starting point, replicates it in a numerical environment, and then further investigates proppant transport through intersecting fractures. As such, $\S$\ref{Section: STIM-LAB experiment} is dedicated to describing the STIM-LAB findings in detail.
	
	Experiments are, however, difficult to conduct in a repeatable manner and interrogate for all of the associated physics. Other intersecting fracture experiments are confined to low-viscosity slickwater fluids, commonly used for shale stimulation. In these experiments, the primary proppant transport mechanism is settling and the subsequent formation and transport of dunes, and the fractures are modelled as slots of finite height~\citep{Wen2016,Tong2017,Sahai2019}. Associated numerical simulations use highly-simplified models, treating the proppant and fluid as one continuous phase~\citep{Han2016,Wang2018a,Akhshik2022}.
	
	There also exists a branch of research which studies the partitioning of particles at idealised T-shaped bifurcations, both experimentally~\citep{Bugliarello1964,Ditchfield1996,Roberts2003,Roberts2006,Manoorkar2016,Manoorkar2018} and numerically~\citep{YezazAhmed2011,Manoorkar2016,Manoorkar2018,Wang2022}, with the intent of garnering insights into complex, physically-realistic scenarios like blood and fracture flows. The channels are typically square or rectangular in cross-section, and have widths which are equal or comparable in size. Moreover, particles freely flow through the secondary channel without blockage or mounding, and the quantitative results are generally presented in terms of the fractions of fluid and particles which enter the secondary channel. The partitioning of particles is primarily dependent on the ratio of flow rates into the respective branches~\citep{Ditchfield1996}, however the two are not directly proportional, particularly at higher flow rates in the primary channel~\citep{Ditchfield1996,Roberts2003,Roberts2006}. In general, partitioning is affected by the distribution of particles in the primary channel, which can be influenced by inertial and shear-induced migration, and for this reason is dependent on $\Rey$ and $\phi$ ~\citep{YezazAhmed2011,Manoorkar2016,Manoorkar2018}. Overall, the physical mechanisms for even single-particle bifurcation are complex~\citep{Doyeux2011}, and for further discussion of the parametric effects and phenomena associated with bifurcation in these types of simplified geometries the reader is referred to \citet{Morris2020}.
	
	These complex behaviours necessitate numerical methods which explicitly track each phase, as opposed to single-phase models based on constitutive rheological equations~\citep{Morris2020}. In general, \ac{DNS} can capture all pertinent physics, while being cheaper and more easy to control than experiments. Its computational intensiveness, however, means that simulations are limited to small length and time scales. Therefore, the general philosophy of \ac{DNS} is to develop accurate quantitative data that can be incorporated in larger scale numerical models, such as those used in \ac{HF} simulators. Recently, \ac{DNS} was applied to clogging in planar fractures~\citep{DiVaira2023}, generating parametric screenout maps which can be scaled to \ac{HF} simulators~\citep{DiVaira2021} and which include complex physics like electrostatics.
	
	However, \ac{DNS} has not been applied in this manner to investigate proppant transport in intersecting fractures. The aforementioned bifurcation studies, while numerous, represent only a small subset of all possible cases of particle transport through secondary channels. Specifically, they are useful for estimating the proportion of diverted particles given known flow rates through the primary and secondary channels, but only when particles freely flow through the secondary channel, and when the secondary channel width is the same order of magnitude as the primary channel width. As already described, however, the reality of \ac{HF} is that proppant must both invade \emph{and} be retained in the cleat (which is orders of magnitude smaller than the hydraulic fracture) via some blockage mechanism in order to prop the cleat, while the formation of occlusions leads to mounding. Existing bifurcation studies (apart from the STIM-LAB study) do not account for these realities, and consequently the insights and data which can be translated from them to inform \ac{HF} treatments are limited.
	
	This paper therefore generates novel data --- including fluid leak-off rates, \acp{SVF} of proppant retained in cleats, and mound sizes --- which account for the realities of proppant placement in intersecting hydraulic and natural fractures, and can therefore immediately inform the design of future proppant treatments. This is achieved with a coupled \ac{LBM}-\ac{DEM} numerical framework, which represents significantly improved parametric control, repeatability, and opportunities for extension in comparison to experiments, ultimately at reduced development and running costs. Firstly, the STIM-LAB experimental setup, key findings and limitations are described in $\S$\ref{Section: STIM-LAB experiment}, followed by a brief description and validation of the \ac{LBM}-\ac{DEM} for shear-thinning fluids in $\S$\ref{Section: Numerical model}. In $\S$\ref{Section: Simplified simulations} the STIM-LAB experiment is then replicated within a simplified computational geometry, generating new qualitative insights and demonstrating the capability of the numerical framework to reproduce the pertinent physical phenomena. Finally, in $\S$\ref{Section: Realistic simulations}, the framework is applied to a more realistic computational environment which comprises rough fracture surfaces, from which the aforementioned novel data are produced.

	\section{Summary of the STIM-LAB experiment}
	\label{Section: STIM-LAB experiment}
	
	The STIM-LAB intersecting fracture experiment, which formed part of the larger coordinated studies in support of \ac{HF} in coals~\citep{Stimlab1995}, had four primary goals, for the ultimate purpose of improving reservoir productivity.
	\begin{enumerate}
		\item Minimise fluid leak-off through the coal cleat during the \ac{HF} stage via the injection of proppant.
		\item Minimise pressure loss due to proppant mounding in the hydraulic fracture.
		\item Maximise cleat regained permeability and proppant pack conductivity.
		\item Investigate the impact of breakers on proppant clean-up and leak-off.
	\end{enumerate}
	However, the scope of this paper is limited to goals one and two. That is, only the proppant transport stage is modelled, with the fluid flowback, cleanup and the resulting productivity neglected. Regarding goal 1, the leak-off of fluid into the reservoir is undesirable because it results in the loss of pumping pressure, which reduces the penetration length and width of the hydraulic fracture, and also increases formation damage due to the chemical compounds present in the fracturing fluid.
	
	\subsection{Setup and parameters}
	
\begin{figure}[h!]
		\centering
		\includegraphics[width=0.5\textwidth]{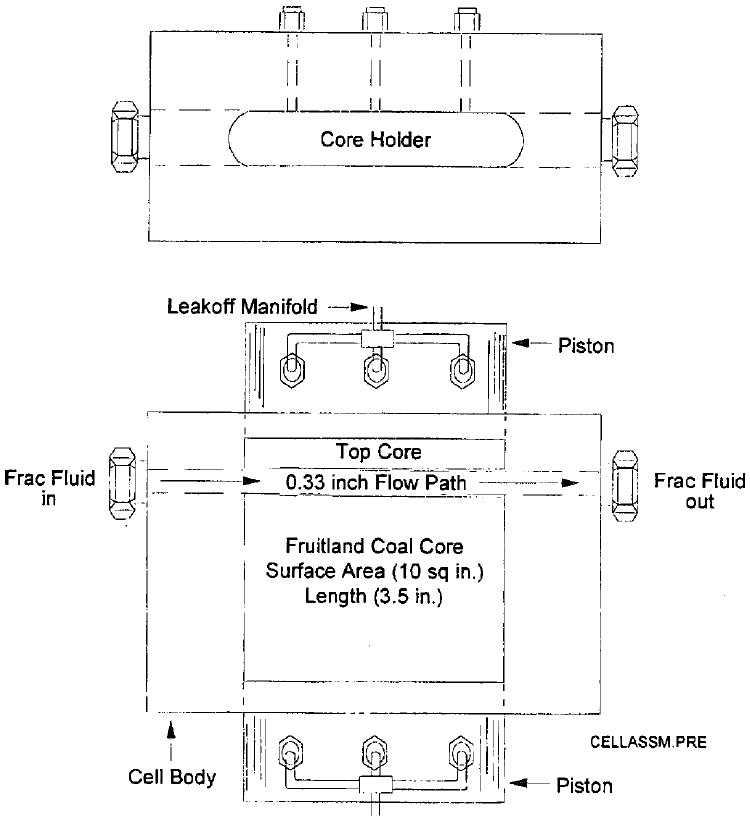}
		\caption{Schematic of the STIM-LAB test cell for investigating leak-off through cleats which intersect a hydraulic fracture, reproduced from the original report~\citep{Stimlab1995}.}
		\label{fig_stimlab_schematic}
	\end{figure}

	A schematic of the experimental test cell is illustrated in Figure~\ref{fig_stimlab_schematic}. It comprises two opposing cores. The bottom core is a cleated coal sample, taken from a representative coal mine, while the top core is continuous coal. The gap between the two samples represents the primary hydraulic fracture, and over the length of the flow path there are therefore one or more cleats on a single side which are perpendicularly aligned to the flow path. A suspension is flowed through the gap and fluid leaks off through the cleats in the bottom core, through which proppant may also pass depending on its size.
	
	This setup allowed the influence of a number of parameters on fluid leak-off to be determined. Firstly, there was a natural variance in cleat sizes between different samples of the same coal. By testing a number of different samples, a number of different cleat widths were therefore also tested. These sizes fell into the range of \SIrange{50}{250}{\micro\metre}. Secondly, three different proppant sizes were tested. These were 100 mesh (\SI{149}{\micro\metre}), 40/70 mesh (\SIrange{210}{425}{\micro\metre}) with 5\% 70/100 mesh (\SIrange{149}{210}{\micro\metre}), and 100/635 mesh (\SIrange{20}{149}{\micro\metre}). However, these sizes are nominal, and in reality the size distributions are much wider. For example, 100 mesh proppant can range from \SIrange{105}{210}{\micro\metre}~\citep{Barree2019}. This is factored into the design of the numerical model in $\S$\ref{Section: Simplified simulations} and $\S$\ref{Section: Realistic simulations}. Finally, the impact of breaker additives on leak-off, which resulted in different fluid characteristics, was also investigated. This aspect of the study is neglected here, however, with the fluid restricted to the standard \SI{35}{\pound} guar plus \SI{1.2}{\pound} borate which was used throughout most of the study. This fluid was moderately shear thinning with $n=0.38$ and $K=$ \SI{0.4}{\pound\per\foot\second\tothe{$n$-2}} (\SI{0.6}{\pascal\second\tothe{$n$}}).
	
	Suspensions were mixed with \SI{2}{\pound} of proppant per gallon of fluid, equating to a bulk \ac{SVF} of $\bar{\phi}=0.1$, and pumped through the fracture at a constant rate of \SI{0.33}{\gallon\per\minute} (\SI{6.31e-5}{\metre\cubed\per\second}).

	\subsection{Findings}
	
	The primary quantified variable of the STIM-LAB study was the leak-off volume versus time. In general the tests followed this pattern: an initial period of high leak-off rate, followed by a period of decrease in leak-off rate as proppant gradually blocked off the cleat, and finally a steady state, where the maximum amount of proppant had built up at the cleat entrance and the leak-off rate decreased with the square-root of time as filter cake built up.
	
\begin{figure}[h!]
		\centering
		\subfloat[]{%
			\includegraphics[width=0.4\textwidth]{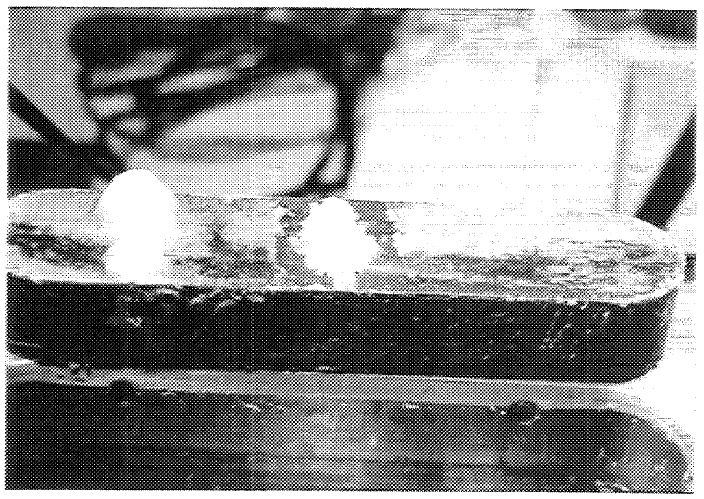}
			\label{fig_stimlab_mounding_100mesh}
		}
		\hspace{1em}
		\subfloat[]{%
			\includegraphics[width=0.4\textwidth]{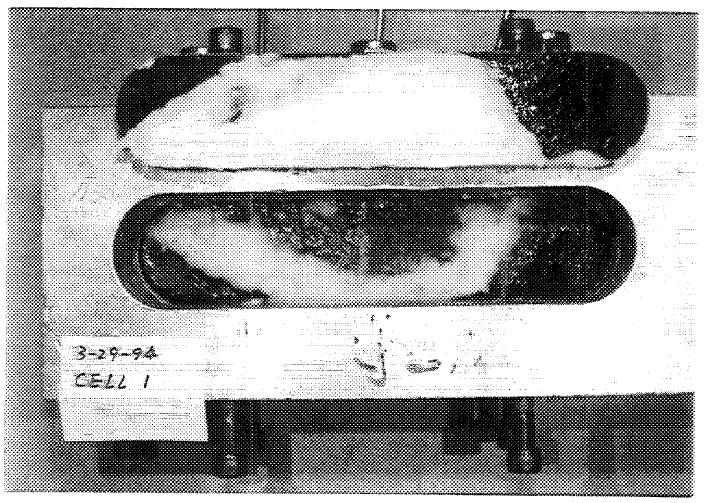}
			\label{fig_stimlab_mounding_4070mesh}
		}
		\hfill
		\subfloat[]{%
			\includegraphics[width=0.4\textwidth]{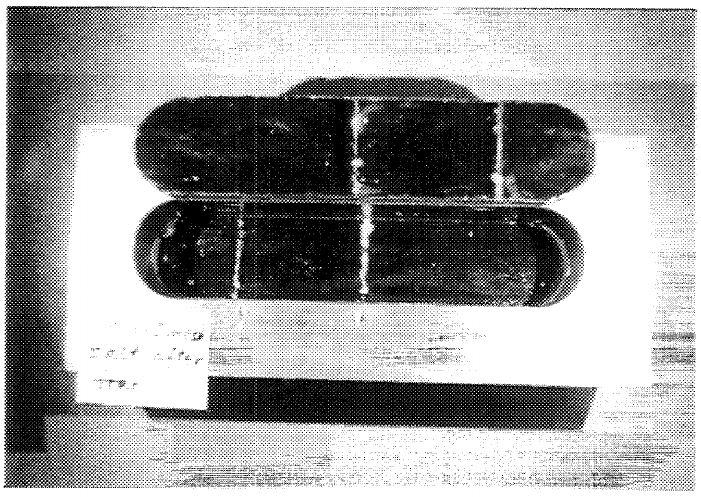}
			\label{fig_stimlab_mounding_100625mesh}
		}
		\caption[Proppant mounding at cleat entrances from the STIM-LAB report]{Proppant mounding at cleat entrances, reproduced from the STIM-LAB report~\citep{Stimlab1995}, for (a) 100 mesh (b) 40/70 mesh and (c) 100/635 mesh.}
		\label{fig_stimlab_mounding}
	\end{figure}

	The study was conducted by investigating the three different proppant size distributions in turn, varying the cleat sizes for each. Beginning with 100 mesh, the cleats were varied in size from \SIrange{40}{160}{\micro\metre}. Firstly, less leak-off occurred for larger cleats. Through examination of the post-test core sample it was discovered that proppant had invaded the larger cleats, and this was therefore postulated as the reason for reduced leak-off. 
	Secondly, for all cleat sizes an occlusion was created at the cleat opening by proppants which were larger than the cleat opening. This occlusion led to significant mounding, as depicted in Figure~\ref{fig_stimlab_mounding_100mesh}, which contributed to a large pressure drop across the hydraulic fracture.
	
	Next, for the 40/70 mesh (with 5\% 70/100 mesh), cleat sizes of \SI{50}{\micro\metre} and \SI{180}{\micro\metre} were tested only. 
	Leak-off was significantly greater for the smaller cleats compared to the larger cleats. As for the 100 mesh, this was the case because the proppant was able to invade the larger cleat, however for the 40/70 mesh the leak-off was much higher. The resulting mounds were also significantly larger, resulting in blockage of most of the primary channel, as shown in Figure~\ref{fig_stimlab_mounding_4070mesh}.
	
	The 100/635 mesh was tested with cleat sizes of \SI{120}{\micro\metre} and \SI{180}{\micro\metre}. 
	For both of these cleat sizes the leak-off was lower than the larger proppants, and similar for both sizes. There was also minimal mounding, as pictured in Figure~\ref{fig_stimlab_mounding_100625mesh}. When the cores were analysed post test, it was found that the cleats were completely filled with proppant. It was also noted, however, that 635 mesh alone flowed straight through the cleat, and consequently did not stop leak-off.
	
	While out of the scope of the numerical simulations subsequently presented in $\S$\ref{Section: Simplified simulations} and $\S$\ref{Section: Realistic simulations}, it is insightful to note the regained permeability achieved with each proppant size. Following each of the leak-off tests, the sodium chloride-based proppants were cleaned up with a combination of larger proppants and a potassium chloride solution. The regained permeability was then measured and found to be highest for 100/635 mesh, and lowest for the 40/70 mesh, confirming that fluid leak-off is detrimental to resulting cleat permeability (for the types of gelled fluids commonly used in \ac{HF} treatments).
	
	In summary, the STIM-LAB findings can be distilled into four qualitative conclusions.
	
	\begin{enumerate}
		\item Larger cleats result in lower fluid leak-off because particles are able to invade the cleats.
		\item Significant mounding occurs at occlusions for 100 mesh and 40/70 mesh proppant.
		\item Leak-off is minimised when the particle size distribution is large (e.g., 100/635 mesh is better than 100 mesh).
		\item Mounding is minimised with 100/635 mesh because leak-off is minimised.
	\end{enumerate}

	\subsection{Shortcomings and gaps}
	
	To close this discussion, it is important to note a number of shortcomings and inconsistencies of the STIM-LAB study. Firstly, cleat sizes were not measured exactly, and were given instead as ranges of sizes combined with a pre-treatment permeability. Secondly, it is unclear how many cleats there were per coal core, and how far apart these were spaced. Next, the variation of parameters was not systematic, nor were the findings presented in a systematic manner. For example, exact cleat sizes reported in the text frequently do not align with those of the referenced figures. Similarly, sodium chloride-based proppants with a salt-saturated fluid were used for some tests, while sand was used for others. Finally, the resulting steady state leak-off rates were constant across all test cases, irrespective of the proppant size and cleat blockage. Intuitively, however, the leak-off flow rate should vary depending on the degree of cleat blockage, which is indeed the case for the numerical results presented next.
	
	For all of these reasons, exact numerical reproduction of the experimental leak-off volumes is impossible. Additionally, the numerical simulations conducted here do not capture the steady-state reduction in leak-off rate due to filter cake build up in the proppant pack and cleat. While the leak-off volumes, and other measurable quantities, will be interpreted quantitatively in the following numerical simulations, comparison to the STIM-LAB findings will be done in a qualitative manner only, e.g., analysing the trends of how leak-off volumes vary between proppant sizes. 
	
	A significant gap, in addition to the above shortcomings, is that no physical explanation for the findings has been given. Furthermore, the exact impact of the shear-thinning fluid rheology, as opposed to a Newtonian fluid, is as yet unknown.

	\section{Rheological and numerical model}
	\label{Section: Numerical model}
	
	The fluid used in the STIM-LAB intersecting fracture experiment, like most hydraulic fracturing fluids, was shear-thinning. Shear-thinning fluids are a type of generalised Newtonian fluid,
	\begin{equation}
		\hat{\tau} = \eta_\text{app}(\dot{\gamma})\dot{\gamma},
		\label{eqn_generalised_newtonian}
	\end{equation}
	where the local shear stress, $\hat{\tau}$, is related to the local shear rate, $\dot{\gamma}$, by a variable apparent viscosity, $\eta_\text{app}$. Numerous types of generalised Newtonian fluids exist, all of which can be described by a rheological equation for $\eta_\text{app}$. One of The most simple is the power-law fluid,
	\begin{equation}
		\eta_\text{app} = K | \dot{\gamma} | ^{\hspace{0.5ex}n-1},
		\label{eqn_power_law}
	\end{equation}
	where $n$ is the dimensionless flow behaviour index and $K$ is the flow consistency index (with units of \SI{}{\pascal\second\tothe{$n$}}, or \SI{}{\pound\per\foot\second\tothe{$n$-2}}). Power-law fluids with $n<1$ exhibit shear-thinning behaviour, i.e., decreasing apparent viscosity with increasing shear rate. In pressure-driven channel flows this manifests in a blunting of the velocity profile relative to the Newtonian Poiseuille profile and an effectively plugged region of zero shear rate at the channel centre. Conversely, shear-thickening power-law fluids, where $n>1$, exhibit an increasing apparent viscosity with increasing shear rate, resulting in a sharpening of the velocity profile.
	
	To model the flow of particles in shear thinning fluids this work employs a coupled \ac{LBM}-\ac{DEM} framework, where the \ac{LBM} models the suspending fluid, the \ac{DEM} integrates the particle movement, and full hydrodynamic coupling between the fluid and particles is achieved using the \ac{PSM}~\citep{Noble1998}. The accuracy of fluid-solid momentum exchange for the \ac{PSM} is well established~\citep{Strack2007, Owen2011}, and for Newtonian fluids the present implementation of the \ac{LBM}-\ac{DEM}-\ac{PSM} has previously been verified and validated for its second-order convergence, bulk suspension behaviour in channel flows (for monodisperse and polydisperse particle size distributions)~\citep{DiVaira2022}, and the short-range hydrodynamics of particle collisions~\citep{DiVaira2023}. The model has also been extensively described in prior works for Newtonian fluids~\citep{DiVaira2022,DiVaira2023}. The \ac{DEM} simulation is performed in the open-source solver LIGGGHTS~\citep{Kloss2012}, while the \ac{LBM} simulation and coupling to \ac{DEM} is performed using the open-source \ac{LBM} code-base, TCLB~\citep{Laniewski2016}.
	
	For shear-thinning fluids, however, an extended \ac{LBM} model which can handle shear rate-dependent viscosity is applied. This shear rate-dependent \ac{LBM} calculates $\dot{\gamma}$ at each node, followed by $\eta_\text{app}$ according to Equation~\eqref{eqn_power_law}. The local apparent viscosity is then set by modifying the local \ac{LBM} relaxation rate. Modelling locally-varying viscosity in this manner leads to two numerical difficulties, which must be appropriately treated. Firstly, by inspection of Equation~\eqref{eqn_power_law}, for values of $n<1$, $\eta_\text{app} \rightarrow \infty$ as $\dot{\gamma} \rightarrow 0$. To overcome this singularity, regularisation \citep{Papanastasiou1987} is used to modify the rheological model,
	\begin{equation}
		\eta_\text{app} = K\dot{\gamma}^{\hspace{0.5ex}n-1}\left(1-e^{-m\dot{\gamma}}\right),
		\label{eqn_regularisation}
	\end{equation}
	where $m$ is a regularisation parameter. To further reduce the numerical instability associated with low shear rates, a linear approximation for $\eta_\text{app}$ is taken from $\eta_\text{app}(\dot{\gamma}=\dot{\gamma}_c)$ to $\eta_\text{app}(\dot{\gamma}=0)$, where $\dot{\gamma}_c$ is the lower shear rate cutoff. In this work values of $\dot{\gamma}_c=\SI{5e-18}{}$ and $m=\SI{1e14}{}$ are used, which are within typical ranges~\citep{Hill2021}. For full details of the shear rate-dependent \ac{LBM} used in this work the reader is referred to~\citet{Leonardi2011}~and~\citet{Hill2021}.
	
	Secondly, when used with just a single relaxation time, the \ac{LBM} solution varies with the choice of the relaxation time, posing a problem to shear rate-dependent formulations. This work therefore uses the \ac{TRT} collision operator~\citep{Ginzburg2008}, which eliminates the solution dependence on the relaxation time.

	\subsection{Validation}
	\label{Section: Numerical model, Subsection: Validation}
	
	The \ac{LBM}-\ac{DEM}-\ac{PSM}, in conjunction with the shear rate-dependent \ac{LBM}, is computationally straightforward to implement. However, novel behaviours arise for particles suspended in flows of shear-thinning fluids, which must therefore be accurately captured by the current numerical model. Furthermore, while the \ac{PSM} is second-order exact and accurate for curved boundaries in Newtonian fluids~\citep{Strack2007}, and \ac{TRT} eliminates the viscosity dependence of the shear rate-dependent \ac{LBM}~\citep{Ginzburg2008}, the integration of the two is not common in the literature, and its numerical accuracy and stability cannot be assumed. This section therefore verifies and validates that the shear rate-dependent \ac{LBM}, in conjunction with the \ac{TRT}, \ac{DEM} and \ac{PSM}, accurately captures the physics of dense particle suspensions in shear-thinning flows.
	
	The validation programme begins with the benchmark case of a shear-thinning fluid flowing past a single, stationary sphere in a tube. Comparison here is made with the literature results of \citet{Song2011}, which were obtained numerically via the \ac{FEM}. For all tests the sphere is positioned on the axis in the flow direction (i.e., at the centre of the tube) and the ratio of sphere diameter to tube diameter is equal to 0.5. Two degrees of shear-thinning are simulated, $n=0.2$, 0.4, as well as the Newtonian case ($n=1$), each at three different particle Reynolds numbers, where the drag coefficient is the measured parameter. To facilitate comparison with the literature study, a particle Reynolds number is defined which accounts for power-law fluids,
	\begin{equation}
		\hat{\Rey}_p = \frac{\rho \langle u \rangle^{2-n} d^n}{K},
		\label{eqn_particle_reynolds_number_power_law}
	\end{equation}
	where the characteristic velocity is the mean channel velocity, $\langle u \rangle$, due to the sphere being fixed, the sphere diameter, $d$, is used as the characteristic size, and $\rho$ is the fluid density. The drag coefficient,
	\begin{equation}
		C_d = \frac{8F_d}{\pi \rho \langle u \rangle^2 d^2},
		\label{eqn_particle_drag_coefficient}
	\end{equation}
	is calculated from the measured drag force on the sphere, $F_d$. For Equations~\eqref{eqn_particle_reynolds_number_power_law}~and~\eqref{eqn_particle_drag_coefficient} the mean channel velocity is calculated analytically from the set maximum channel velocity, $u_\text{max}$, and power-law index,
	\begin{equation}
		\langle u \rangle = u_\text{max} \left( \frac{n+1}{3n+1} \right).
	\end{equation}
	For the present simulations, the length of the domain is set to ten particle diameters in order to allow the flow to revert to its pure-fluid profile after passing the sphere. A high computational resolution of $d/\delta_x=25$ is used, where $\delta_x$ is the \ac{LBM} lattice spacing, noting that the errors associated with lattice resolution are investigated next.
	
\begin{figure}[h!]
		\hspace{-2.4em}\centering\includegraphics{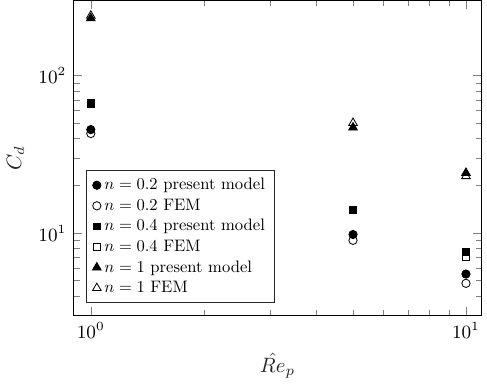}
		\caption{Variation of the drag coefficient, $C_d$, with the particle Reynolds number for power-law fluids, $\hat{\Rey}_p$, for the benchmark case of a fixed sphere in a tube as predicted by the present model implementation for shear rate-dependent fluids. Comparison is made to literature results obtained with the \ac{FEM}~\citep{Song2011} for increasing degrees of shear-thinning, represented by the flow behaviour index, $n$.}
		\label{fig_validation_powerlaw_sphere}
	\end{figure}

	The results in Figure~\ref{fig_validation_powerlaw_sphere} demonstrate that the present model correctly produces the general trends of decreasing drag with both increasing Reynolds number and decreasing flow behaviour index. Firstly, matching of the case of $n=1$ shows that the shear rate-dependent \ac{LBM} accurately models Newtonian fluids, i.e., where the viscosity remains constant throughout the domain. For the shear-thinning cases, drag is evidently well predicted by the present model, with small errors introduced at high $\hat{\Rey}_p$ only.
	
\begin{figure}[h!]
		\hspace{-2em}\centering\includegraphics{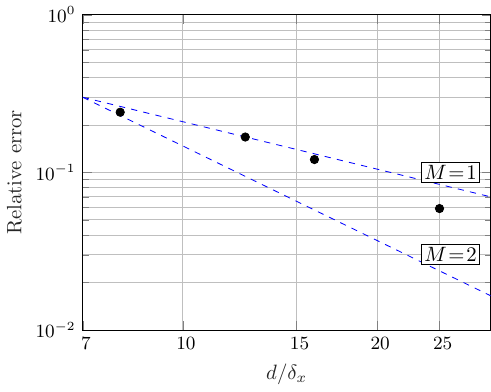}
		\caption{Numerical convergence of the present shear rate-dependent \ac{LBM}-\ac{PSM} implementation (using \ac{TRT}), obtained via the fixed sphere in a tube benchmark case at $n=0.2$, $\hat{\Rey}_p=10$. The relative error is calculated in comparison to the \ac{FEM} result~\citep{Song2011}, where $M$ is the slope of convergence.}
		\label{fig_convergence_powerlaw}
	\end{figure}

	To analyse the numerical accuracy of the shear rate-dependent \ac{LBM}-\ac{PSM} implementation, the solution for the case of $n=0.2$ at $\hat{\Rey}_p=10$ is obtained for lattice resolutions over the range of $d/\delta_x=8$ to 25. The error at each resolution is calculated relative to the \ac{FEM} solution and plotted in Figure~\ref{fig_convergence_powerlaw}, demonstrating numerical convergence. It is noted that, when the same test is run with the single relaxation time \ac{LBM} (as opposed to \ac{TRT}), no convergence occurs.
	
	Finally, the present shear rate-dependent model is extended from single spheres to dense suspensions. Specifically, it is validated for the case of shear-induced migration of a monodisperse suspension in a shear-thinning fluid, against results obtained with the \ac{DFM}~\citep{Reddy2019}. Shear-induced migration is the phenomenon of bulk particle diffusion in the direction of diminishing shear rate. For channel flows of Newtonian fluids it results in an accumulation of particles at the channel centre and a blunting of the velocity profile. Here, the same test cell from the work of \citet{DiVaira2022}, which investigated the shear-induced migration of polydisperse suspensions in a Newtonian fluid, is used. In the present case the channel has dimensions of $l \times w \times h = 8.6d \times 23.2d \times 5.7d$, with wall particles of size $d_w=d/1.05$, and a computational resolution of $d/\delta_x=11.2$. Simulations are run until the velocity profile reaches full development. The test case here comprises a moderately dense suspension, $\bar{\phi}=0.3$, of a moderately shear-thinning base fluid, $n=0.5$. At this moderate \ac{SVF}, the shear-thinning fluid should cause further blunting (relative to that already caused by shear-induced migration for the Newtonian case) of the velocity profile to the point of flattening, as predicted by the DFM.
	
	\begin{figure}[h!]
		\hspace{-2em}\centering\includegraphics{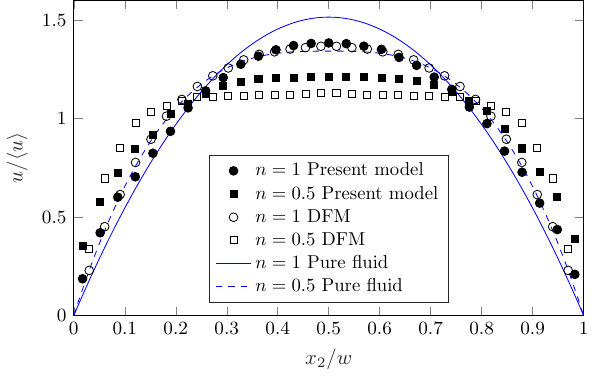}
		\caption{Validation of shear-induced migration for a moderately dense ($\bar{\phi}=0.3$) suspension with a moderately shear-thinning ($n=0.5$) base fluid. Comparison is made to the \ac{DFM}~\citep{Reddy2019} and the Newtonian case ($n=1$).}
		\label{fig_migration_powerlaw}
	\end{figure}
	
	As expected, Figure~\ref{fig_migration_powerlaw} demonstrates close agreement between the present model and the \ac{DFM} for the Newtonian case ($n=1$), where the peak has been blunted relative to the pure fluid. For the shear-thinning case ($n=0.5$), the present model correctly predicts the complete flattening of the velocity profile, but not to the same degree as the \ac{DFM}. This discrepancy between the two numerical methods likely arises due to the \ac{DFM}'s use of an apparent viscosity correction based on local solid volume fraction. However, to the best of the authors' knowledge this is the only literature data available and, for the purpose of the subsequent work in this paper, this matching of the correct general behaviour is deemed to suffice.

	\section{Simplified simulations}
	\label{Section: Simplified simulations}
	
	Here a simplified computational geometry is developed, where the hydraulic fracture has a reduced size and the cleat is represented as a smooth, tapered channel. The aim is to replicate the proppant transport aspect of the STIM-LAB study, and in doing so demonstrate the capability of the numerical framework to reproduce the pertinent physical phenomena. Subsequently, in $\S$\ref{Section: Realistic simulations}, a physically-realistic computational environment is developed and exact quantitative data is obtained.
	
	\subsection{Simplified test cell}
	\label{Section: Simplified simulations, Subsection: Simplified test cell}
	
\begin{figure}[h!]
		\centering
		\subfloat[]{
			\includegraphics[width=0.9\textwidth]{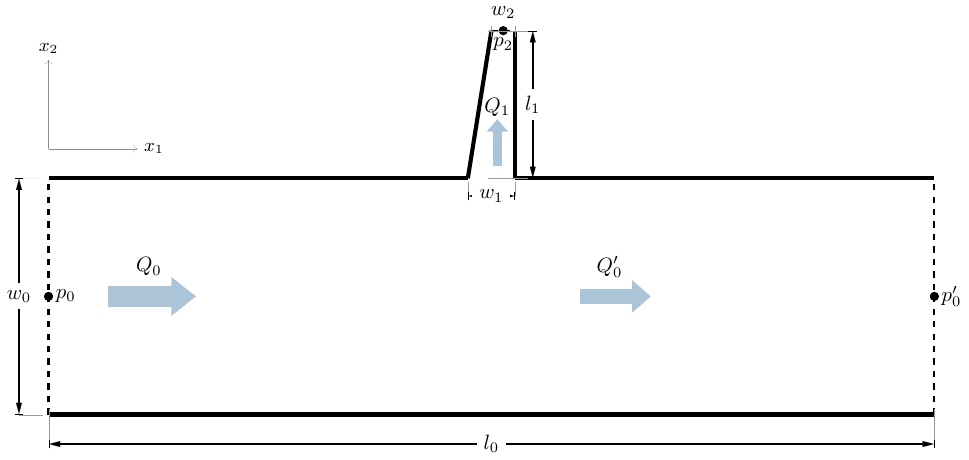}
			\label{fig_test_cell_simplified_schematic}
		}
		
		\vspace{1em}
		\subfloat[]{
			\includegraphics[width=0.7\textwidth]{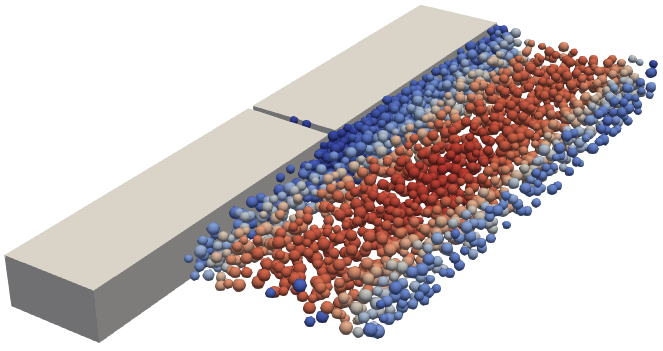}
			\label{fig_test_cell_simplified_3d}
		}
		\caption{The simplified numerical test cell designed to validate the numerical model for proppant flow through intersecting fractures, showing (a) a two-dimensional schematic, where solid lines represent solid walls, dotted lines denote constant-pressure boundaries, and the simulation is periodic in the $x_3$ direction (into the page). The flow is mass conserving, such that $Q_0=Q_0'+Q_1$, and the cleat is drawn at a larger scale for clarity of illustration. A three-dimensional perspective of 100 mesh proppant flowing past and mounding at a \SI{160}{\micro\metre} cleat is shown in (b), where the wall which opposes the cleat is omitted for clarity.}
	\end{figure}

	There are two key considerations for replicating the physical experiment in a numerical simulation. The first relates to representing the variability of the cleat surface, which determines the local variations in the cleat width. In $\S$\ref{Section: Realistic simulations} the cleat surfaces are represented with realistic surface roughness by a statistical model, and the sensitivity of the results to the associated statistical parameters is determined. Here, however, for the purpose of demonstrating the validity of the numerical model, a simplified numerical test cell is devised which represents the cleat as a tapering channel, as illustrated in Figure~\ref{fig_test_cell_simplified_schematic}. This tapering channel is a deterministic representation of the varying channel width of a realistic cleat, where the width linearly decreases along the channel. This also ensures that larger proppant can be retained in the cleat once it enters, while smaller proppant (such as 635 mesh) is allowed to flow straight through, which replicates the physical reality. The tapered channel is completely represented by three parameters: the inlet (maximum) width, $w_1$; the outlet (minimum) width, $w_2$; and the channel length, $l_1$. Throughout the results in this section, $w_1$ is the quoted nominal width of the cleats. A consistent taper is used such that $w_1 - w_2 = \SI{60}{\micro\metre}$, and $l_1=\SI{1.5}{\milli\metre}$. To further illustrate the numerical setup and better convey the nature of the simulations, Figure~\ref{fig_test_cell_simplified_3d} captures the case of 100 mesh flowing past a \SI{160}{\micro\metre} cleat from a three-dimensional perspective.
	
	The second issue is that the physical length scales of the primary hydraulic fracture width and the cleat width can differ by a factor of $10^3$. The spatial resolution required to resolve the cleat via \ac{DNS} therefore means that also simulating the entire hydraulic fracture becomes intractable. Instead, the approach taken in this paper is to simulate only a truncated section of the primary hydraulic fracture, in both the flow and transverse directions, seeing as the entrance to the cleat is the primary region of interest. For the simplified test cell, the flow is bounded by a solid wall on the side of the fracture opposing the cleat (for the realistic test cell, the flow is approximated with a linear velocity gradient, as described in further detail in $\S$\ref{Section: Realistic simulations}). The simulation is periodic in the neutral direction, with a height of \SI{760}{\micro\metre}, and the length and width of the fracture channel are $l_0=\SI{10}{\milli\metre}$ and $w_0=\SI{2.76}{\milli\metre}$, respectively. 
	
	Finally, the time scale of the numerical simulations is necessarily smaller than the physical experiment (seconds compared to minutes) due to computational constraints (the time step must be sufficiently small to satisfy numerical stability, and decreases with the square of the spatial resolution according to diffusive scaling). Therefore, the flow rates and mound sizes are scaled down compared to the physical experiment. The desired net fluid fluxes are achieved by applying constant pressure boundaries at the inlet and two outlets of the domain. Doing so allows the flow rates to decrease according to the degree of particle build-up in the test cell, as opposed to directly specifying the flux at each boundary, which would maintain a constant flow rate throughout the simulation. The selection of these pressure boundary values is discussed next.

	\subsection{Simulation parameters}
	\label{Section: Simplified simulations, Subsection: Simulation parameters}
	
	The parameters which are varied in the simplified simulations are the cleat width, the proppant size, and the fluid type, as presented in Table~\ref{tab_leakoff_parameters}. Firstly, to replicate the inter-dependencies between these parameters reported in the STIM-LAB experiment, it is sufficient to test only two different cleat widths.
	
	\begin{table}[h!]
		\centering
		\small
		\caption{The parameters varied in the simplified intersecting fracture simulations.}
		\begin{tabular}{ll}
			\toprule
			Cleat widths ($w_1$) & \begin{tabular}{l} \SI{160}{\micro\metre} \\ \SI{240}{\micro\metre} \end{tabular}\\
			\midrule
			Proppant sizes 	& \begin{tabular}{l} 100 mesh \\ 40/70 mesh \\ 100/635 mesh \end{tabular}\\
			\midrule
			Fluid types 	& \begin{tabular}{l} Newtonian \\ Shear-thinning \end{tabular}\\
			\bottomrule
		\end{tabular}
		\label{tab_leakoff_parameters}
	\end{table}
	
	Secondly, both Newtonian and shear-thinning fluids are implemented to determine their comparative influences on leak-off. The Newtonian fluids are assigned a physical kinematic viscosity of $\nu=\SI{1e-6}{\metre\squared\per\second}$, while the shear-thinning fluid has $n=0.38$ and $K =$ \SI{0.6}{\pascal\second\tothe{$n$}} (\SI{0.4}{\pound\per\foot\second\tothe{$n$-2}}) to match the experimental fluid reported in the STIM-LAB experiment. Newtonian fluids are simulated with the standard \ac{LBM}-\ac{PSM}, while shear-thinning fluids are simulated with the shear rate-dependent \ac{LBM}-\ac{PSM}. It is noted that the shear rate-dependent model requires a time step which is 6.667 times smaller than the Newtonian model in order to maintain stability.
	
	To maintain consistency between the different tests of different parametric combinations in these simplified simulations, the ratio of the flow rates at the primary fracture inlet to the cleat for a pure fluid simulation (without particles), $Q_{f,0}/Q_{f,1}$, is held constant. The pressures at each boundary required to achieve this are therefore initially determined in a fluid-only simulation for each combination of cleat width and fluid type. These pressures are subsequently used for all different particle sizes at that combination. The pressures and flow rates for each parameter combination are given in Table~\ref{tab_leakoff_flow_rates}. Seeing as the pressure gradients of the STIM-LAB experiment are not known, a direct comparison to the numerical pressures cannot be made. Even if they were known, there would be little correspondence, seeing as the scales of the fracture width and cleat length differ by orders of magnitude.
	
	Particles are constantly injected with a random injection seed in the region $x_1=$ \SIrange{1.5}{2}{\milli\metre}, such that a \ac{SVF} of $\bar{\phi}=0.1$ is maintained.
	
	\begin{table}[h!]
		\centering
		\small
		\caption{Selection of pressure boundary values for the simplified simulations for each parameter combination using a pure-fluid simulation, such that the ratios of flow rates at the primary fracture inlet and cleat, $Q_{f,0}/Q_{f,1}$, are consistent. Higher flow rates are possible for the shear-thinning case due to the smaller time step.}
		\begin{tabular}{lcccccc}
			\toprule
			Fluid & $p_0 - p_0'$ (\SI{}{\pascal}) & $p_0 - p_2$ (\SI{}{\pascal}) & $w_1$ (\SI{}{\micro\metre}) & $Q_{f,0}$ (\SI{}{\metre\cubed\per\second}) & $Q_{f,1}$ (\SI{}{\metre\cubed\per\second}) & $Q_{f,0}/Q_{f,1}$ \\
			\midrule
			Newtonian & 0.021 & 3.021 & 160 & \SI{5.04e-9}{} & \SI{0.38e-9}{} & 13.46 \\
			Newtonian & 0.021 & 0.841 & 240 & \SI{5.04e-9}{} & \SI{0.38e-9}{} & 13.46 \\
			Shear-thinning & 15 & 215 & 160 & \SI{15.30e-9}{} & \SI{1.15e-9}{} & 13.27 \\
			Shear-thinning & 15 & 102.5 & 240 & \SI{15.30e-9}{} & \SI{1.15e-9}{} & 13.27 \\
			\bottomrule
		\end{tabular}
		\label{tab_leakoff_flow_rates}
	\end{table}

	\subsection{Proppant size distributions}
	\label{Section: Simplified simulations, Subsection: Proppant size distributions}
	
	As described in $\S$\ref{Section: STIM-LAB experiment}, the proppant sizes reported in the STIM-LAB experiment are nominal only, and in reality have wide distributions. Determining how the sizes should be distributed, as well as the maximum and minimum sizes at which the distribution should be truncated, are therefore key considerations for numerical simulations. Here these choices for the simplified simulations are described, along with their simplifying assumptions. In $\S$\ref{Section: Realistic simulations} the influence of the \ac{PSD} is analysed in more depth.
	
\begin{figure}[h!]
		\centering\includegraphics[width=0.82\linewidth]{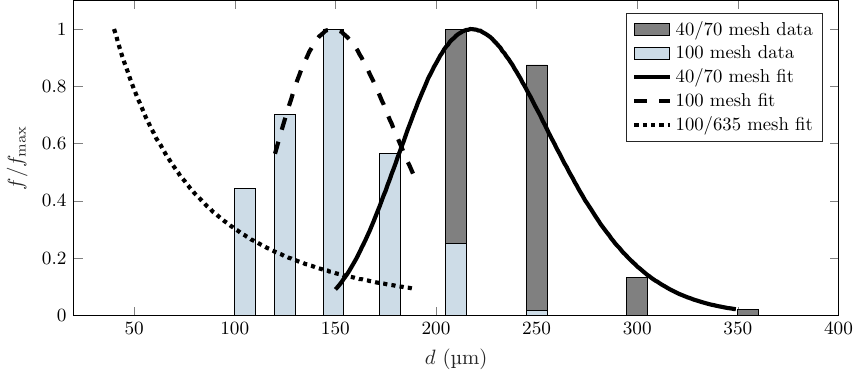}
		\caption{Proppant size distributions from a literature analysis~\citep{Barree2019} (for 40/70 mesh and 100 mesh), along with the normalised probability distribution functions of their log-normal fits, $f/f_\text{max}$, utilised in the simplified numerical simulations. Fits are plotted only between the minimum and maximum sizes at which they are truncated.}
		\label{fig_proppant_distributions}
	\end{figure}

	Representing a \ac{PSD} with a mathematical function is difficult, seeing as there is no standard specification for how the sizes should be distributed (mainly due to the difficulty and cost of size control at the micron scale). The first requirement is therefore to select an appropriate probability distribution function. In this paper a log-normal distribution
	is used for all distributions, seeing as it best represents naturally-occurring sands and synthesised ceramics~\citep{Wagner1994}. 
	Secondly, seeing as the proppant \acp{PSD} are not analysed in the STIM-LAB report, data from a size distribution analysis in the literature~\citep{Barree2019} are used instead for the 100 mesh and 40/70 mesh (with 5\% 70/100 mesh). These data are plotted in Figure~\ref{fig_proppant_distributions}, along with the truncated log-normal fits which are implemented in the simplified simulations. Seeing as 100/635 mesh, however, is composed of two different proppant types (100 mesh and 635 mesh silica flour), its distribution is somewhat arbitrary. Here, therefore, a log-normal distribution which represents particles in the range \SIrange{40}{190}{\micro\metre} is fit.
	
	The maximum and minimum truncation sizes for each distribution are selected based on two factors. Firstly, the distribution must be sufficiently wide so that occlusions occur at the cleat entrance due to size exclusion of the largest particles. On the other hand, the size of the smallest particle is limited by the computational resolution, which in turn determines the size of the largest particles (to ensure the distribution is not lopsided). It is for this second reason that the 100 mesh is truncated to such a large extent, and that the 100/635 mesh is truncated at \SI{40}{\micro\metre}, even though 635 mesh is nominally \SI{20}{\micro\metre}. The 40/70 mesh is extended to include smaller sizes than the literature data so that the fit includes some 100mesh, recalling that the STIM-LAB 40/70 mesh included 5\% 70/100 mesh. The influence of distribution truncation is investigated in detail in $\S$\ref{Section: Realistic simulations}.
	
	In terms of the lattice resolutions, for the 100 mesh and 40/70 mesh a lattice spacing of \SI{20}{\micro\metre} is used, resulting in minimum resolutions of $d_\text{min}/\delta_x=6$ and 7.5, respectively, while for the 100/635 mesh a lattice spacing of \SI{10}{\micro\metre} is used, resulting in a minimum resolution of $d_\text{min}/\delta_x=4$. Finally, the particle sizes are implemented in increments of \SI{10}{\micro\metre} in the \ac{DEM} to closely approximate a continuous distribution.

	\subsection{Results and analysis of simplified simulations}
	\label{Section: Simplified simulations, Subsection: Results}
	
\begin{figure}[h!]
		\centering
		\small
		\stackunder[0.7em]{\includegraphics[width=0.48\textwidth]{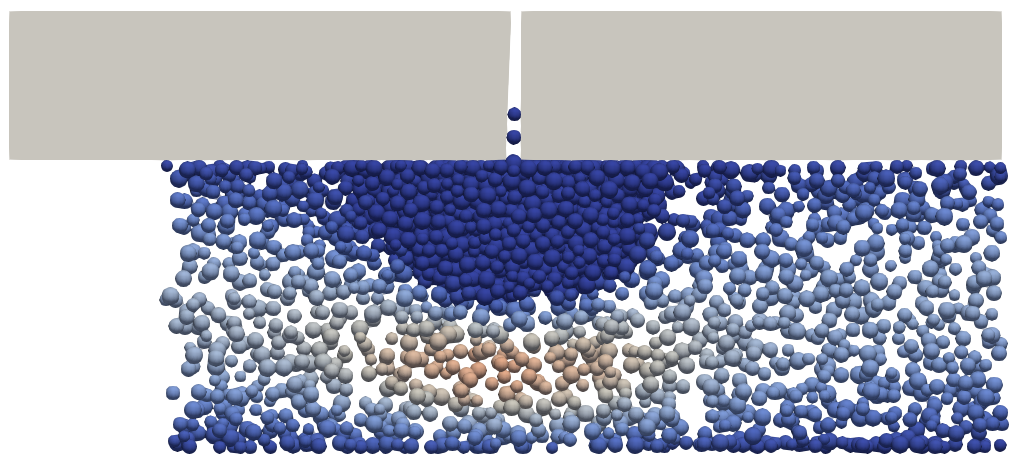}}{100 mesh, \SI{160}{\micro\metre} cleat}
		\hspace{0.6em}
		\stackunder[0.7em]{\includegraphics[width=0.48\textwidth]{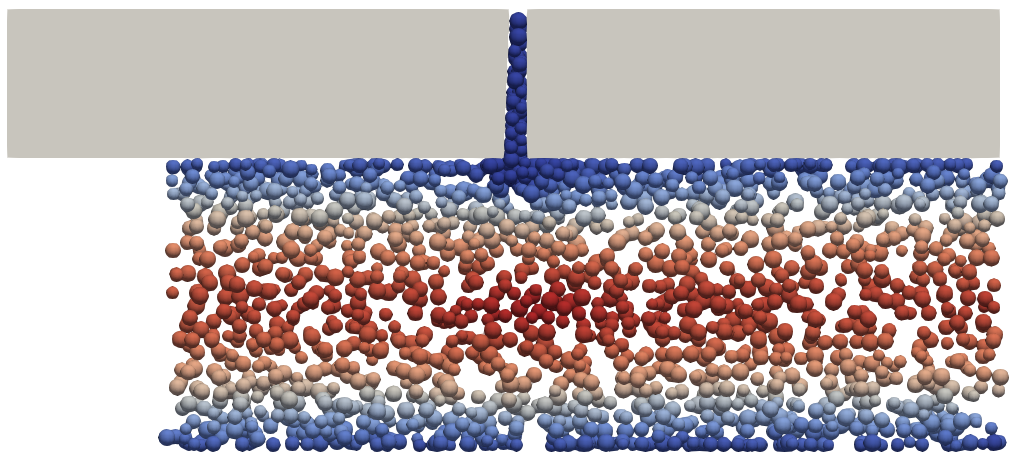}}{100 mesh, \SI{240}{\micro\metre} cleat}
		\newline
		
		\vspace{1.5em}
		\stackunder[0.7em]{\includegraphics[width=0.48\textwidth]{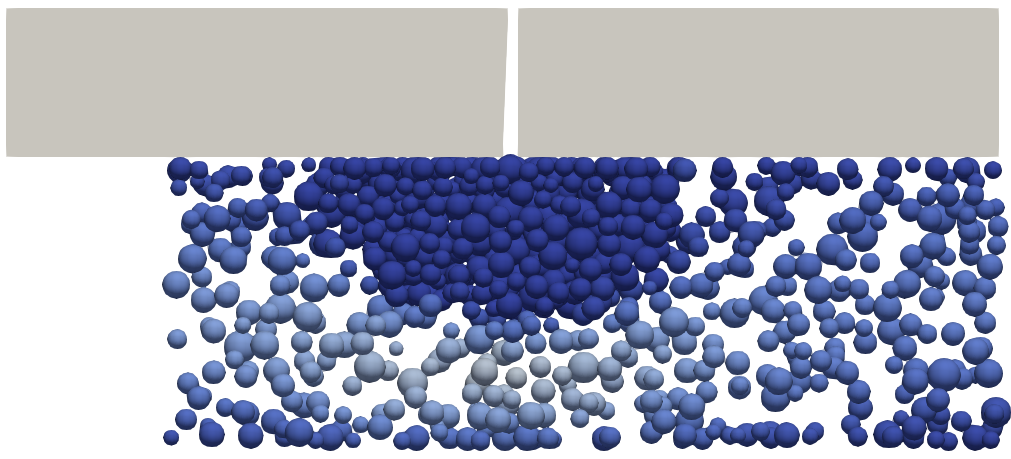}}{40/70 mesh, \SI{160}{\micro\metre} cleat}
		\hspace{0.6em}
		\stackunder[0.7em]{\includegraphics[width=0.48\textwidth]{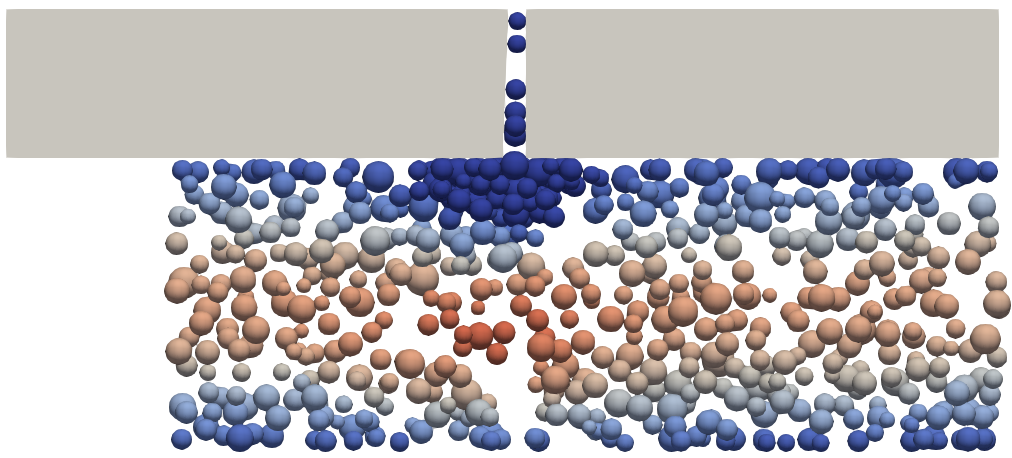}}{40/70 mesh, \SI{240}{\micro\metre} cleat}
		\newline
		
		\vspace{1.5em}
		\stackunder[0.7em]{\includegraphics[width=0.48\textwidth]{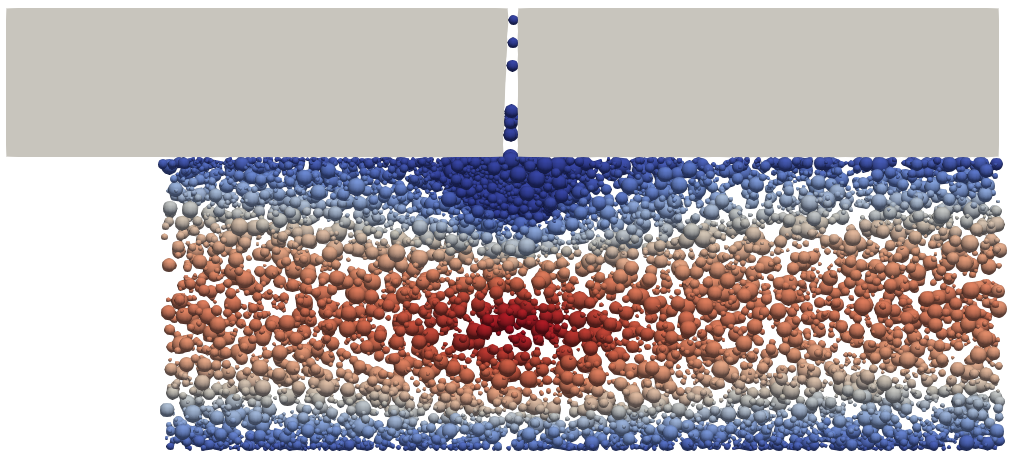}}{100/635 mesh, \SI{160}{\micro\metre} cleat}
		\hspace{0.6em}
		\stackunder[0.7em]{\includegraphics[width=0.48\textwidth]{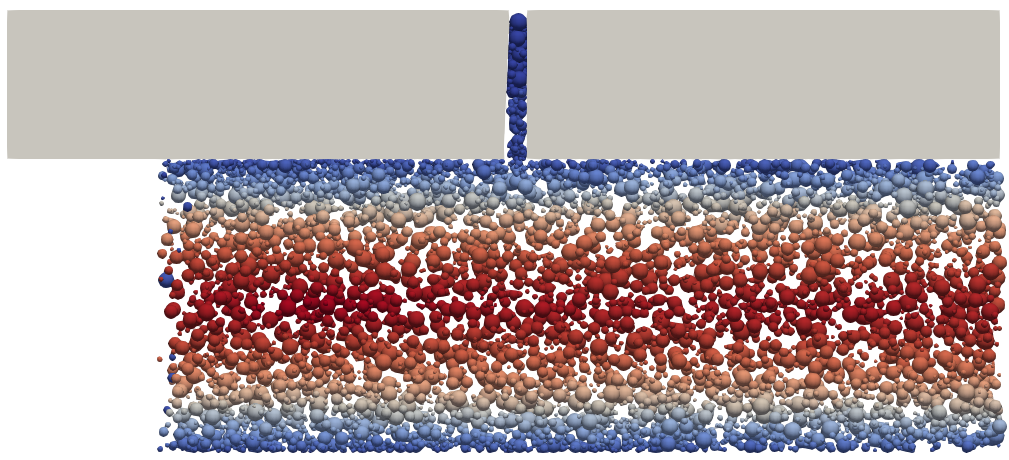}}{100/635 mesh, \SI{240}{\micro\metre} cleat}
		\newline
		
		\vspace{2em}
		\stackunder[0.4em]{\includegraphics[width=0.35\textwidth, height=5mm]{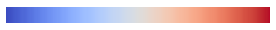}}{$| \boldsymbol{u}_p |$}
		
		\caption{Steady-state distribution of particles through the main fracture and cleat of the simplified simulations, for the Newtonian fluid at various proppant distributions and cleat sizes. Images are captured from a two-dimensional top view. Particles are coloured by their velocity magnitude, $| \boldsymbol{u}_p |$, using the same scale for all images.}
		\label{fig_leak_screenshots_newt}
	\end{figure}

	The simplified simulations are firstly conducted for the Newtonian fluid only, for the cleat sizes and proppant distributions given in Table~\ref{tab_leakoff_parameters}. Figure~\ref{fig_leak_screenshots_newt} depicts the resulting distribution of particles through the primary fracture and cleat at steady-state conditions, demonstrating that the mounding observed in the STIM-LAB experiments has been reproduced. For the \SI{160}{\micro\metre} cleat the largest mounding occurs for the 40/70 mesh and 100 mesh, while the 100/635 mesh is significantly smaller. This is consistent with conclusions 2 and 4 of the STIM-LAB findings. For the larger \SI{240}{\micro\metre} cleat mounding has been reduced for the 40/70 mesh and 100 mesh, and eliminated altogether for the 100/635 mesh. It is also evident that the velocity in the primary channel reduces significantly in cases of large mounding.
	
\begin{figure}[h!]
		\centering
		\subfloat[]{%
			\hspace{-2.4em}\includegraphics{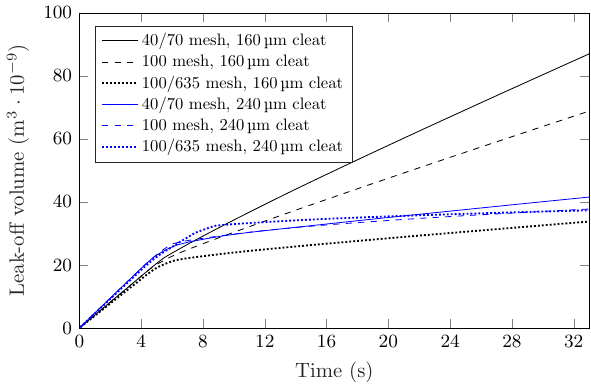}
			\label{fig_leak_volume_newt}
		}
		
		\subfloat[]{%
			\hspace{-3.2em}\includegraphics{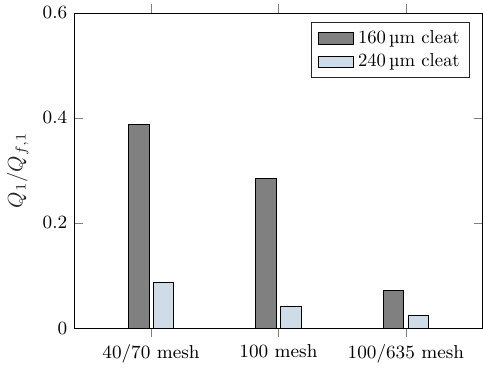}
			\label{fig_leak_rates_newt}
		}
		
		\caption{Leak-off volumes and rates for the simplified simulations using a Newtonian fluid and all combinations of cleat and proppant sizes, where (a) is the cumulative total of fluid which has leaked through the cleat over time and (b) is the leak-off rate at steady state conditions, $Q_1$, normalised by the pure-fluid leak-off rate, $Q_{f,1}$.}
		\label{fig_leak_plots_newt}
	\end{figure}

	To investigate these findings in further detail, the leak-off volumes and rates are analysed. Figure~\ref{fig_leak_volume_newt} firstly plots the increases in leak-off volumes with time, showing that the leak-off reaches a steady state after an initial period of high leak-off rate, as particles gradually block and or fill the cleats. Figure~\ref{fig_leak_rates_newt} then plots the steady-state rates. Firstly, the smaller cleat results in significantly greater leak-off than the larger cleat and, referring to Figure~\ref{fig_leak_screenshots_newt}, this coincides with the larger cleat being invaded by the particles, which is consistent with conclusion 1 of the STIM-LAB findings. Secondly, the 100/635 mesh has significantly reduced the leak-off in both cleats, which is consistent with conclusion 3. The fact that this reduced leak-off coincides with the reduced mounding corroborates conclusion 4 of the STIM-LAB findings, i.e., that mounding is minimised due to leak-off being minimised.
	
\begin{figure}[h!]
		\centering
		\stackunder[0.7em]{\includegraphics[scale=0.9]{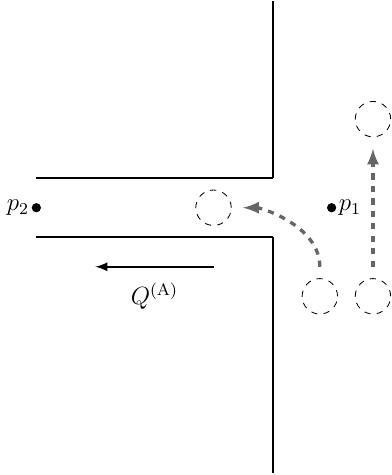}}{Scenario A}
		\hspace{5em}
		\stackunder[0.7em]{\includegraphics[scale=0.9]{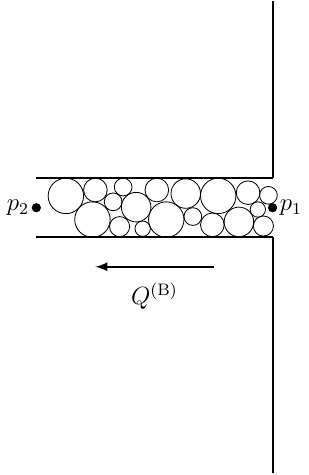}}{Scenario B}
		\newline
		
		\vspace{2em}
		\stackunder[0.7em]{\includegraphics[scale=0.9]{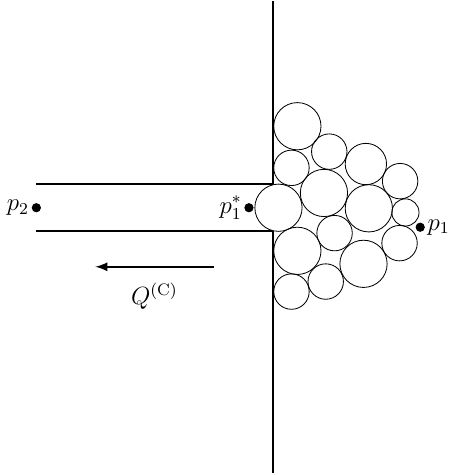}}{Scenario C}
		\vspace{1.em}
		
		\caption{Three scenarios which help to explain mounding and flow rates through an intersecting cleat, where the comparative flow rates are related as $Q^{\text{(A)}} > Q^{\text{(C)}} > Q^{\text{(B)}}$. Here $p_2$ is the outlet pressure, which is the same in all scenarios, and $p_1$ is the pressure required to overcome the flow in the main channel and pull a particle into the cleat, which must therefore necessarily increase further from the cleat entrance. Scenario A represents a pure fluid; in Scenario B an additional pressure drop has been created which moves $p_1$ to the cleat entrance; and in Scenario C the pressure at the cleat entrance, $p_1^{*}$, is sufficient to attract particles and cause mounding.}
		\label{fig_leak_scenarios}
	\end{figure}

	To provide a physical explanation for these findings, the Kozeny-Carman equation~\citep{Kozeny1927,Carman1937},
	\begin{equation}
		\frac{\Delta p}{l} = \frac{150 \eta_f}{\Phi_s^2 d^2} \frac{\left( 1 - \epsilon \right)^2}{\epsilon^3} \bar{u},
		\label{eqn_KC}
	\end{equation}
	can be used, which describes the pressure drop, $\Delta p$, through a packed particle bed of length $l$, given a superficial velocity, $\bar{u}$. By taking the sphericity, $\Phi_s$, to be equal to one for perfect spheres, converting the porosity, $\epsilon$, to the \ac{SVF}, $\phi$, and replacing $\bar{u}$ with the flow rate by dividing by the superficial area, $A$, Equation~\eqref{eqn_KC} can be reformulated,
	\begin{equation}
		\frac{\Delta p}{l} = \frac{150 \eta_f}{d^2 A} \frac{\phi^2}{\left( 1 - \phi \right)^3} Q.
		\label{eqn_KC_simplified}
	\end{equation}
	This can be contrasted with the pressure drop in a planar channel without particles,
	\begin{equation}
		\frac{\Delta p}{l} = \frac{12 \eta_f}{w^2} \bar{u}.
		\label{eqn_pressure_drop_channel}
	\end{equation}
	In conjunction with these equations, three different scenarios are imagined, as illustrated in Figure~\ref{fig_leak_scenarios}, to help explain the flow rates and mounding seen in the results. 
	
	Firstly, it must be recognised that the same pressure drop occurs along the length of the main channel for all scenarios because its inlet and outlet are at constant pressures. Therefore, at a point which is in line with the cleat, but sufficiently far from the cleat entrance, the pressure must always be the same. Point 1, and its corresponding pressure, $p_1$, can then be imagined as the equilibrium point at which the pressure from the cleat is sufficiently low to overcome the drag force in the main channel and pull a particle into the cleat. For a pure fluid (scenario A), this occurs somewhere just outside the cleat entrance. As particles fill the cleat and $\phi$ increases in the cleat, an additional pressure drop is created according to Equation~\eqref{eqn_KC_simplified}. Scenario B represents the case where this pressure drop is just large enough to move the equilibrium point to the cleat entrance, meaning that particles will flow past the cleat without attaching. This nearly corresponds to the 100 mesh in the \SI{240}{\micro\metre} cleat in Figure~\ref{fig_leak_screenshots_newt}. For the 100/635 mesh in the \SI{240}{\micro\metre} cleat, sufficient pressure drop has occurred prior to the cleat being completely filled, due to the higher $\phi$ achieved by the mixture of small and large particles. 
	
	To explain the decrease in cleat flow rate from scenario A to B, it can be physically intuited that the presence of particles has reduced the effective permeability through the cleat. Mathematically, by equating Equations~\eqref{eqn_KC_simplified}~and~\eqref{eqn_pressure_drop_channel}, it is also straightforward to see that, for the same pressure drop, a particle-filled channel must have a significantly lower $Q$ than a pure-fluid channel.
	
	The explanation of mounding (scenario C), and why $Q^{\text{(C)}} > Q^{\text{(B)}}$, is more complex. In scenario C an occlusion occurs at the cleat entrance due to the size exclusion of large particles. In this case minimal to no particles penetrate the cleat, resulting in only a small pressure drop through the cleat. The resulting pressure at the cleat entrance, $p_1^{*}$, will then be large enough to attract particles, leading to mounding. The mound will continue to grow in size, creating an increasingly larger pressure drop, until the pressure at the edge of the mound reaches the critical value of $p_1$ where particles will no longer be added. Theoretically, insufficient pressure drop could occur when the cleat is completely filled with particles, which would also lead to mounding, however this is not observed.
	
	The mound can then be viewed as a porous bed which has much larger flow area ($A$) than the cleat. Consequently, seeing as the flow area is greater for scenario C compared to scenario B but $\left( p_1 - p_2 \right)$ is the same, the flow rate must also be greater for scenario C according to Equation~\eqref{eqn_KC_simplified}.

	\subsubsection{Shear-thinning comparison}
	
\begin{figure}[h!]
		\centering
		\small
		\stackunder[0.7em]{\includegraphics[width=0.48\textwidth]{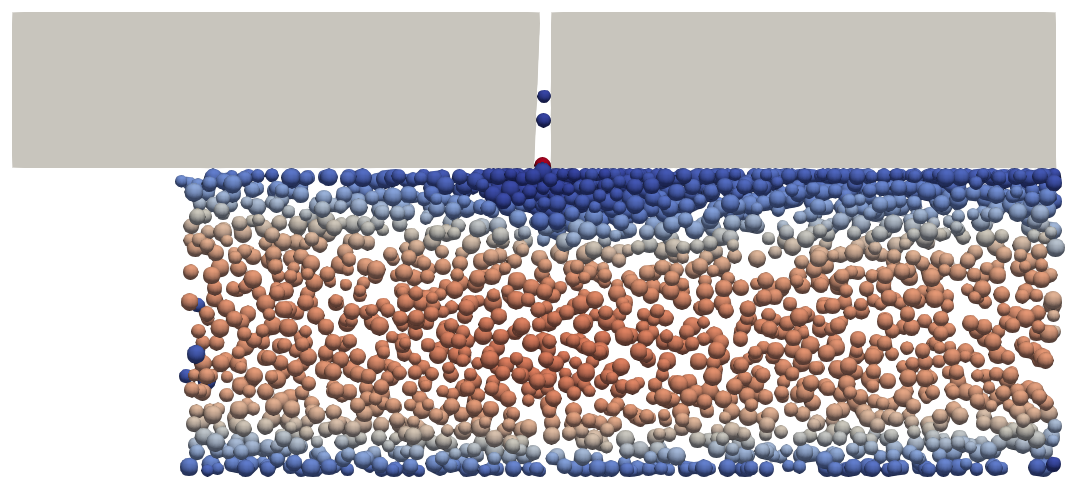}}{100 mesh, \SI{160}{\micro\metre} cleat}
		\hspace{0.6em}
		\stackunder[0.7em]{\includegraphics[width=0.48\textwidth]{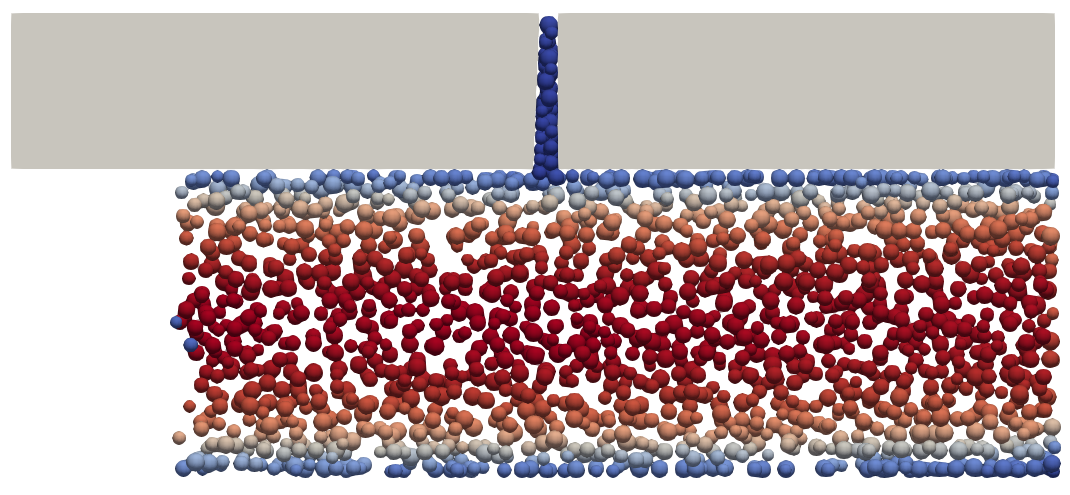}}{100 mesh, \SI{240}{\micro\metre} cleat}
		\newline
		
		\vspace{1.5em}
		\stackunder[0.7em]{\includegraphics[width=0.48\textwidth]{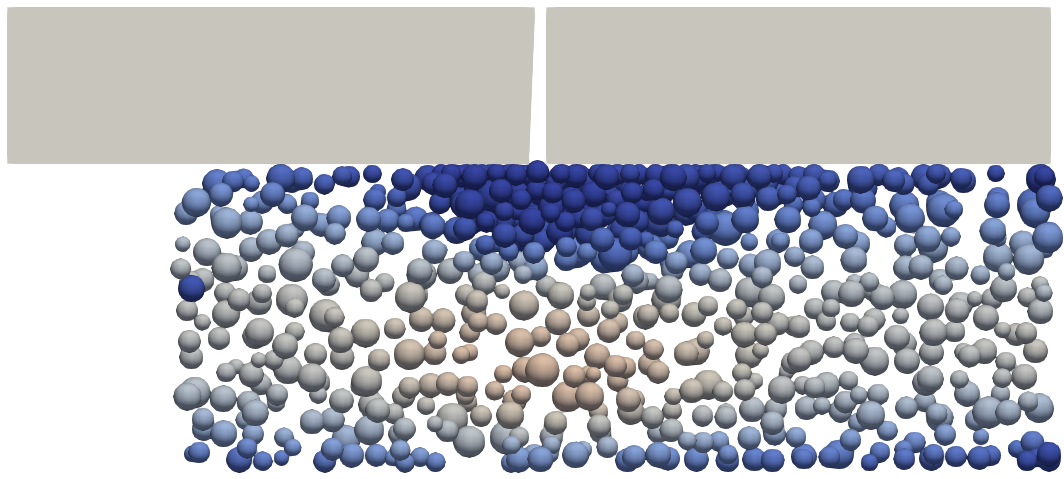}}{40/70 mesh, \SI{160}{\micro\metre} cleat}
		\hspace{0.6em}
		\stackunder[0.7em]{\includegraphics[width=0.48\textwidth]{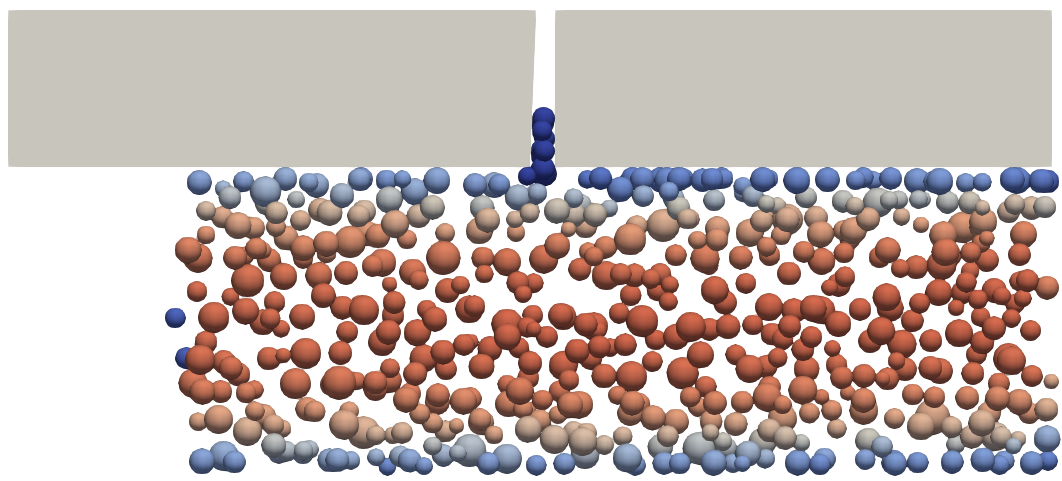}}{40/70 mesh, \SI{240}{\micro\metre} cleat}
		\newline
		
		\vspace{1.5em}
		\stackunder[0.7em]{\includegraphics[width=0.48\textwidth]{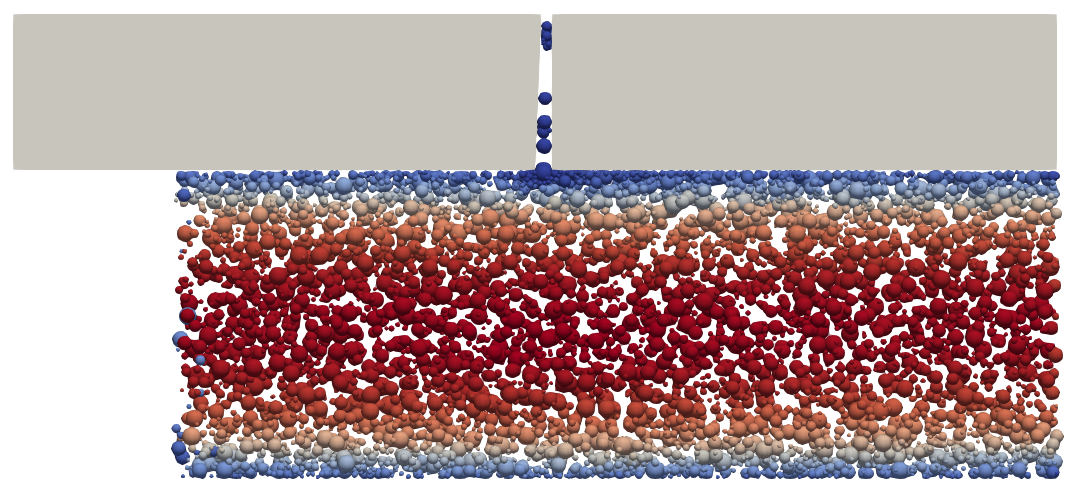}}{100/635 mesh, \SI{160}{\micro\metre} cleat}
		\hspace{0.6em}
		\stackunder[0.7em]{\includegraphics[width=0.48\textwidth]{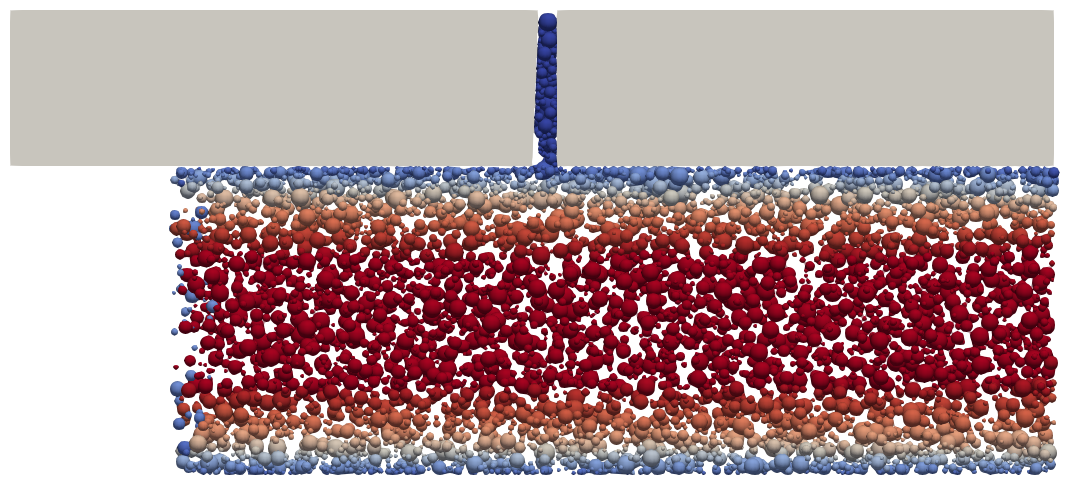}}{100/635 mesh, \SI{240}{\micro\metre} cleat}
		\newline
		
		\vspace{2em}
		\stackunder[0.4em]{\includegraphics[width=0.35\textwidth, height=5mm]{fig_colourmap_redblue.png}}{$| \boldsymbol{u}_p |$}
		
		\caption{Steady-state distribution of particles through the main fracture and cleat of the simplified simulations, for the shear-thinning fluid at various proppant distributions and cleat sizes. Images are captured from a two-dimensional top view. Particles are coloured by their velocity magnitude, $| \boldsymbol{u}_p |$, using the same scale for all images.}
		\label{fig_leak_screenshots_shear}
	\end{figure}

	Here the simplified simulations are repeated, now with the shear-thinning fluid, recalling from $\S$\ref{Section: Simplified simulations, Subsection: Simulation parameters} that the simulations of different fluid types and cleat widths are standardised by matching the pure fluid leak-off ratio, $Q_{f,0}/Q_{f,1}$. Firstly, the resulting distribution of particles through the primary fracture and cleat are depicted in Figure~\ref{fig_leak_screenshots_shear}. Compared to the Newtonian fluid, the mounds have been eliminated in most cases, and have significantly decreased for the 100 mesh and 40/70 mesh in the \SI{160}{\micro\metre} cleat.
	
\begin{figure}[h!]
		\centering
		\subfloat[]{%
			\hspace{-2.4em}\includegraphics[scale=1]{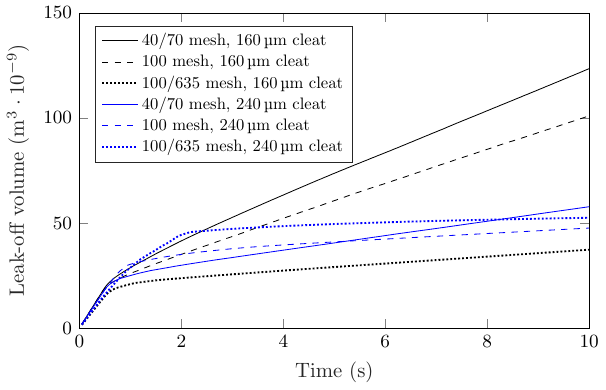}
			\label{fig_leak_volume_shear}
		}
		
		\subfloat[]{%
			\hspace{-2.8em}\includegraphics[scale=1]{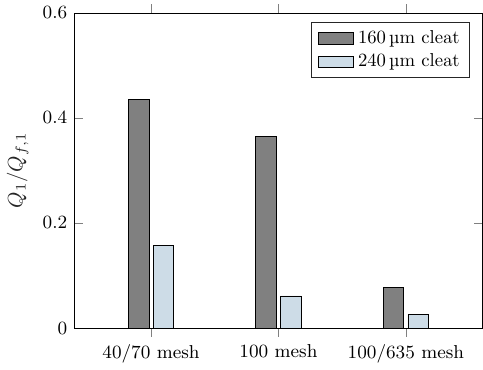}
			\label{fig_leak_rates_shear}
		}
		
		\caption{Leak-off volumes and rates, for the simplified simulations, for the shear-thinning fluid and all combinations of cleat and proppant sizes, where (a) is the cumulative total of fluid which has leaked through the cleat over time and (b) is the leak-off rate at steady state conditions, $Q_1$, normalised by the pure-fluid leak-off rate, $Q_{f,1}$.}
		\label{fig_leak_plots_shear}
	\end{figure}

	The leak-off volumes and rates are plotted in Figure~\ref{fig_leak_volume_shear}. Unlike the mounding, however, these are qualitatively similar to the Newtonian results (Figure~\ref{fig_leak_volume_newt}), in that the smaller cleat results in greater leak-off than the larger cleat, and the 100/635 mesh has reduced leak-off in both cleats. Comparatively, however, the leak-off rates of the shear-thinning fluid are slightly higher.
	
	Overall, it can therefore be concluded that shear-thinning fluids are more desirable for controlling leak-off, seeing as they result in significantly less mounding for similar reductions in leak-off rates. This is due to the fluid viscosity, $\eta_f$, being much higher for a shear-thinning fluid compared to a Newtonian fluid for the low shear rates which occur through a densely-packed bed. By inspection of Equation~\eqref{eqn_KC_simplified}, this higher $\eta_f$ must correspond to a shorter length, $l$ (i.e., the size of the mound), over which the same $\Delta p$ occurs.

	\section{Realistic simulations}
	\label{Section: Realistic simulations}
	
	By using a simplified computational geometry it was demonstrated in $\S$\ref{Section: Simplified simulations} that the present numerical framework is capable of reproducing the qualitative characteristics of physical intersecting fracture experiments. Now, in this section, the same numerical framework is applied to simulations which are more commensurate with physical reality, for the purpose of generating quantitative data which can directly inform future \ac{HF} treatments.
	
	There are three characteristics which make the following simulations more ``realistic''. Firstly, all length scales are physical, due to the unique design of the test cell. Secondly, the cleats are modelled as rough surfaces, which now becomes the mechanism for proppant capture and retention in the cleats. Thirdly, the output parameters are non-dimensional, rendering them transferable to the length and time scales of an entire \ac{HF}. The first two of these characteristics are discussed in further detail when the setup of the test cell is presented in $\S$\ref{Section: Realistic simulations, Subsection: Realistic test cell}, and the third is discussed in the parameterisation in $\S$\ref{Section: Realistic simulations, Subsection: Simulation parameters}. Finally, the results are presented and analysed in $\S$\ref{Section: Realistic simulations, Subsection: Realistic results}.
	
	\subsection{Realistic test cell}
	\label{Section: Realistic simulations, Subsection: Realistic test cell}
	
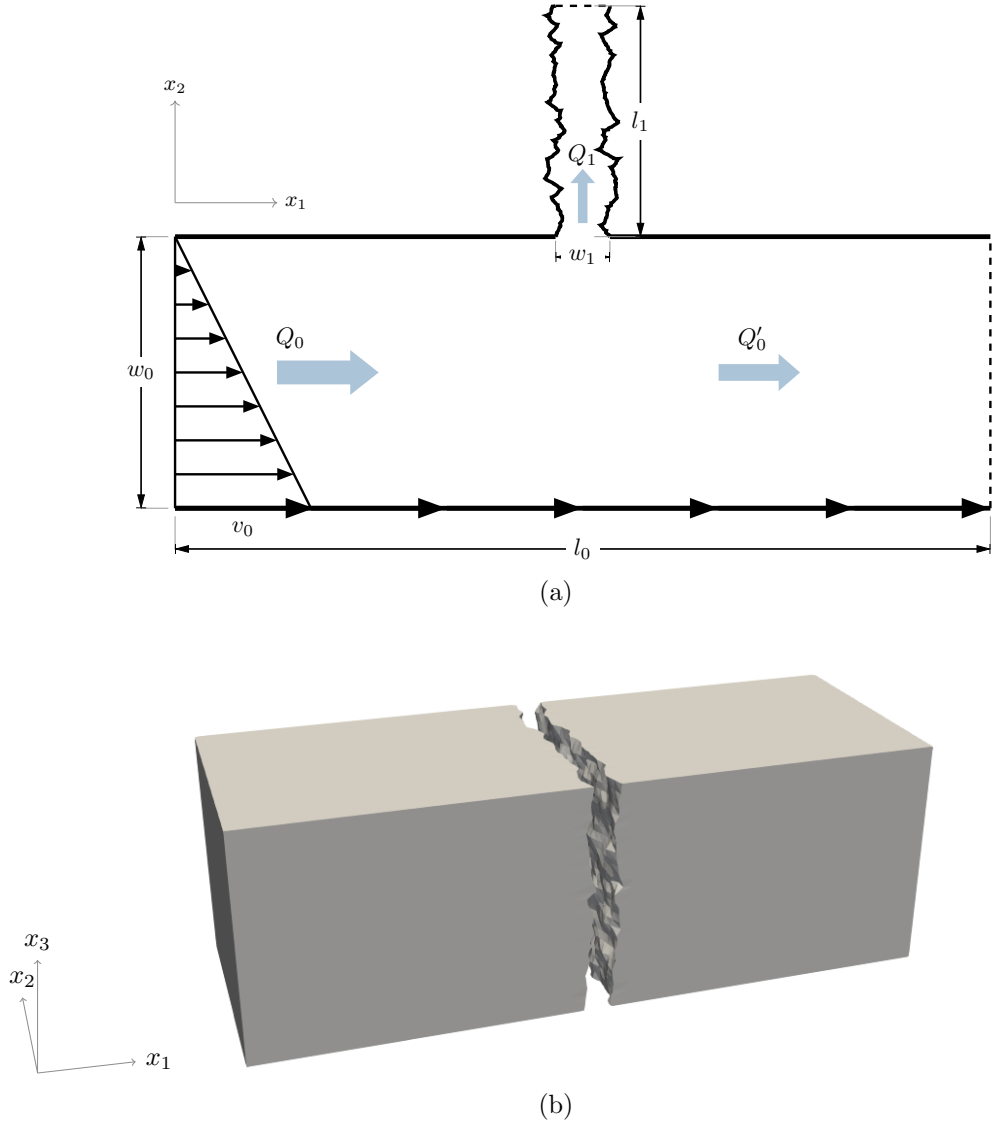
\begin{figure}
		\centering
		\subfloat[]{
			\includegraphics[width=0.7\linewidth]{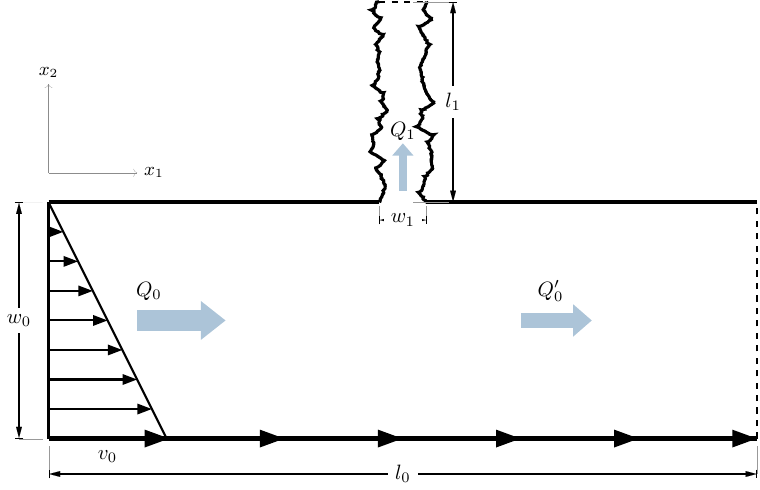}
			\label{fig_test_cell_realistic_schematic}
		}
		
		\vspace{1em}
		\subfloat[]{
			\hspace{-6em}
			\begin{tikzpicture}
				\node[anchor=south west,inner sep=0] at (0,0) {\includegraphics[width=0.6\linewidth]{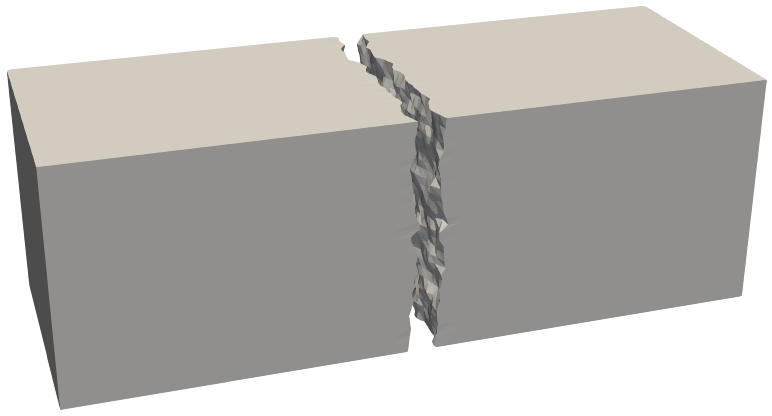}};
				\draw[->, color=gray!90, ultra thin, font=\footnotesize] (-2,0) -- (-0.7,0.15) node[right, color=black]{$x_1$};%
				\draw[->, color=gray!90, ultra thin, font=\footnotesize] (-2,0) -- (-2.2,1) node[above, color=black]{$x_2$};%
				\draw[->, color=gray!90, ultra thin, font=\footnotesize] (-2,0) -- (-2,1.5) node[above, color=black]{$x_3$};%
			\end{tikzpicture}
			\label{fig_test_cell_realistic_3d}	
		}
		\caption{The realistic numerical test cell, with rough cleat, designed to generate the novel data for proppant flow through intersecting fractures, showing (a) a two-dimensional schematic (approximately to scale), highlighting the linear velocity profile at the inlet and the moving wall opposing the cleat, with velocity $v_0$. The dotted lines at the two outlets denote constant-pressure boundaries, the flow is mass conserving, such that $Q_0=Q_0'+Q_1$, and the simulation is periodic in the $x_3$ direction (into the page). A three-dimensional perspective of the solid walls for $w_1=\SI{260}{\micro\metre}$ and $\sigma=\SI{5}{\micro\metre}$ is shown in (b), with the moving wall which opposes the cleat omitted for clarity.}
		\label{fig_test_cell_realistic}
	\end{figure}

	A schematic of the realistic intersecting fracture test cell is depicted in Figure~\ref{fig_test_cell_realistic_schematic}. Its first defining feature is the rough cleat, which is effectively a rough fracture cut from a three-dimensional rectangular block, as illustrated in Figure~\ref{fig_test_cell_realistic_3d}. This rough fracture is generated using the package published at \href{https://github.com/llaniewski/rfracture/}{https://github.com/llaniewski/rfracture/}, a detailed description of which is given in \citet{Laniewski2020}. The two surfaces are realisations of correlated Gaussian random fields described by a power-law spectrum, which have constant means of $\frac{1}{2}w_1$ and $-\frac{1}{2}w_1$, where $w_1$ is the reference cleat width. The standard deviation of the surfaces from their means, $\sigma$, is equivalent to the root mean square of the physical roughness, with units of $\SI{}{\metre}$. In this work the two surfaces are perfectly correlated (i.e., are identical). Therefore, the cleat aperture is constant, and any size exclusion or arching of particles as they travel along the cleat will be only due to tortuosity, the magnitude of which is governed by $\sigma$. In this way, it can be hypothesised that higher cleat surface roughness will lead to greater proppant retention in the cleat, which is indeed demonstrated in the results in $\S$\ref{Section: Realistic simulations, Subsection: Realistic results}. 
	
	The final fracture parameters which are important to note are the decay ratio, $\alpha$, and the rolling frequency, $f_\text{roll}$, of the power spectrum used to generate the Gaussian fields. Combined, these two parameters essentially determine the wavelengths which are cut off when generating the surfaces. Seeing as the cleat lengths modelled in this work are very short, it is desirable to retain only short-wavelength characteristics, i.e., to eliminate any long-wavelength variations which might emerge in longer fractures. To achieve this, values of $\alpha=3.5$ and $f_\text{roll}=\SI{2}{\per\milli\metre}$ are used.
	
	In this work, $\sigma$ and $w_1$ are the only fracture parameters which are varied and, in the results in $\S$\ref{Section: Realistic simulations, Subsection: Realistic results}, detailed maps of different parameters of interest are generated in terms of $\sigma$ and $w_1$. However, seeing as the cleat surfaces are generated from random fields, two cleats with the same $\sigma$ and $w_1$ will be different, and the simulation results will therefore also differ. Therefore, each data point in the parametric maps in the results is the mean of five different realisations at the same $\sigma$ and $w_1$. To capture this randomness with more accuracy, however, an improved stochastic model which utilises many more fracture realisations is needed.
	
	The second defining feature of the realistic test cell is that a small fraction of the total width of the hydraulic fracture, which is orders of magnitude larger than the width of a cleat, is simulated without compromising the physical reality. This is done by applying a linear velocity profile to the flow in the primary channel. This effectively approximates the near-wall region of a much larger channel (i.e., the region which is a sufficiently small fraction of the total width), seeing as, for a parabolic flow, the velocity profile near the wall can be approximated as linear. In the \ac{LBM} simulation, the wall opposing the cleat is given a constant velocity, $v_0$, and a constant linear velocity profile is applied at the inlet, as illustrated in Figure~\ref{fig_test_cell_realistic_schematic}. Overall, this results in a constant shear rate across the hydraulic fracture, $\dot{\gamma}_0=v_0/w_0$, as well as a constant flow rate at the inlet. The two outlets are modelled with pressure boundaries to conserve mass while allowing the flow rates in each channel to vary depending on the degree of cleat blockage. The selection of boundary values, and therefore flow rates, are discussed next in $\S$\ref{Section: Realistic simulations, Subsection: Simulation parameters}.
	
	The hydraulic fracture is approximated as smooth, assuming that the cleat width and roughness are the primary variables of interest. This also minimises the number of parameters which must be investigated. Furthermore, modelling only a single cleat assumes that the cleat spacing is large enough for cleats not to interfere with one another. Finally, it must be ensured that the simulations are independent of the computational domain size. In particular, this relates to the length and height of the cleat, seeing as these two parameters influence the clogging of particles in the cleat, which in turn influences the final results. In a recent work of clogging in planar channels~\citep{DiVaira2023}, it was demonstrated that a height and length of $15d$ and $16d$, respectively, were sufficient to eliminate any domain dependence. Therefore, based on a midrange particle size of \SI{150}{\micro\metre} for the 100 mesh \ac{PSD}, a domain height of $h=\SI{2.25e-3}{\metre}$ and a cleat length of $l_1=\SI{2.4e-3}{\metre}$ are selected for the realistic test cell (for which only 100 mesh proppant is tested). The dimensions of the primary channel are $l_0=\SI{6.6e-3}{\metre}$ and $w_0=\SI{2e-3}{\metre}$, which ensure that the flow at the cleat entrance and any mounding are not affected by the moving wall or the primary channel inlet and outlet.
	
	It is emphasised, however, that the simulated cleat length is orders of magnitude shorter than the cleat network of a reservoir. This implies that, in reality, there is a higher probability of proppant retention. In other words, because proppant travels further along cleats in an actual reservoir compared to these simulations, it has more chance of being captured by size exclusion. This represents a limitation of the realistic test cell.
	
	\subsection{Simulation parameters}
	\label{Section: Realistic simulations, Subsection: Simulation parameters}
	
	In terms of length scales, the numerical lengths exactly match the physical lengths, seeing as a small region of the entire hydraulic fracture is modelled by approximating the velocity gradient as linear. The numerical time scale, however, is necessarily orders of magnitude smaller than the physical time scale, due to the computational cost and stability requirements of the \ac{LBM}, as already discussed in $\S$\ref{Section: Simplified simulations, Subsection: Simplified test cell}. This makes the numerical flow rates and mound sizes also smaller than the physical ones. Therefore, to give them physical meaning for the purpose of informing future \ac{HF} treatments, the flow rates and mound sizes are non-dimensionalised here. The non-dimensional leak-off rate,
	\begin{equation}
		\bar{Q} = \frac{Q_1}{w_1^2h\dot{\gamma}_0},
		\label{eqn_leak_nondim}
	\end{equation}
	effectively converts the leak-off flow rate, $Q_1$, to the shear rate in the cleat ($Q_1/(w_1^2h)$), then divides this by the shear rate in the hydraulic fracture, $\dot{\gamma}_0$. In other words, $\bar{Q}$ is the non-dimensional representation of the shear rate in the cleat versus the shear rate in the fracture channel. Similarly, the non-dimensional mound height,
	\begin{equation}
		\bar{H} = \frac{Hw_1h\dot{\gamma}_0}{Q_1},
		\label{eqn_mound_nondim}
	\end{equation}
	is based on the analysis from $\S$\ref{Section: Simplified simulations} that the mound height, $H$, is proportional to the pressure drop in the cleat, and therefore the superficial cleat velocity ($Q_1/(w_1h)$), according to the Kozeny-Carman equation. Equation~\eqref{eqn_mound_nondim} is therefore the ratio of the mound height to the superficial cleat velocity, also normalised by $\dot{\gamma}_0$ to account for shearing effects in the main channel. $H$ is measured as the distance from the cleat entrance to the centre of the outermost particle in the mound, which is determined as the particle with zero velocity and smallest $x_2$ coordinate value. The final non-dimensional quantity which is of interest is the \ac{SVF} of proppant retained in the cleat, $\phi'$, which is calculated by dividing the volume of stationary particles in the cleat by the nominal volume of the cleat, $l_1w_1h$.
	
	The velocity gradient imposed at the inlet sets $\dot{\gamma}_0$ and the total mass flow rate, $Q_0$, while the pressure boundaries at the exit of the main channel and the cleat determine the respective flow rates through each. However, seeing as the flow rate through the cleat varies with the combination of $w_1$ and $\sigma$, a method of standardising all flow rates is required. This is done by selecting $\dot{\gamma}_0$ and the pressure boundaries so that the ratio of superficial velocity in the cleat to the shear rate in the channel, $Q_1/(w_1h\dot{\gamma}_0)$, is equal to 4.2 for all $w_1$ at $\sigma=0$ (i.e., a smooth cleat). Then, the same boundary values are used for all $\sigma$ at the same $w_1$. The total simulation time for all tests are sufficient for the leak-off rates, retained \ac{SVF} and/or mound sizes to reach a steady state.
	
	The fluid is shear-thinning with $n=0.4$ and $K =$ \SI{0.6}{\pascal\second\tothe{$n$}} (\SI{0.4}{\pound\per\foot\second\tothe{$n$-2}}), closely matching the guar-borate solution used in the STIM-LAB investigation, and generally representative of a \SI{40}{\pound\per\mega\gallon} guar mixture~\citep{Cho2000}. 
	
	A bulk \ac{SVF} of $\bar{\phi}=0.1$ is continually injected at the inlet. However, an additional test is run at a higher \ac{SVF}, $\bar{\phi}=0.4$, which is explained further in $\S$\ref{Section: Realistic simulations, Subsection: Particle size distributions}. The main channel is initially filled with particles to accelerate the process of reaching steady state.

	\subsection{Proppant size distributions and high solid volume fraction}
	\label{Section: Realistic simulations, Subsection: Particle size distributions}
	
	These realistic simulations are conducted for 100 mesh proppant only. The 100 mesh \ac{PSD} from the literature~\citep{Barree2019}, which was used for the simplified simulations (Figure~\ref{fig_proppant_distributions}), is also used here. However, now the distribution is truncated much less --- at minimum and maximum values of \SI{105}{\micro\metre} and \SI{225}{\micro\metre} --- representing nearly the full range of the physical distribution. The effect of truncating the \ac{PSD} is also analysed in the results.
	
\begin{figure}
		\centering\includegraphics{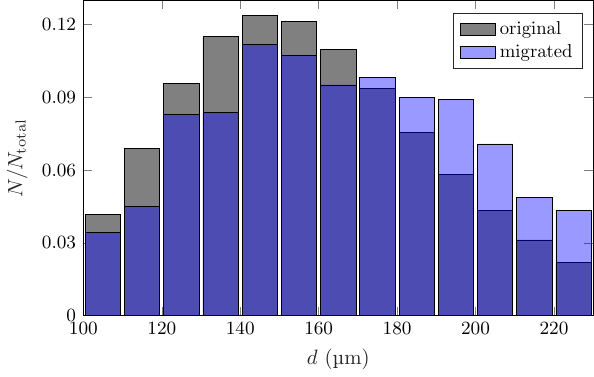}
		\caption{The 100 mesh \acp{PSD} implemented in the realistic simulations. \emph{Grey} is based on the non-truncated measurements from the literature~\citep{Barree2019}, \emph{light blue} is the near-wall region of the original distribution which has migrated in a straight channel at $\bar{\phi} = 0.5$, and \emph{dark blue} is the overlap of the distributions. \acp{PSD} are plotted as number fraction of particles at each size, $N/N_\text{total}$.}
		\label{fig_distributions_high-phi}
	\end{figure}

	In a recent work~\citep{DiVaira2022} it was demonstrated that, for a polydisperse suspension of high \ac{SVF} ($\bar{\phi} \gtrsim 0.4$) in a pressure-driven channel flow, the smallest particles form the plug at the channel centre, while the largest particles migrate to the channel walls. In terms of fracture flows, this has the implication that the near-wall region will comprise a higher proportion of larger particles compared to the entire suspension, which most likely influences proppant and fluid leak-off into the cleats. Therefore, as a way of translating this novel, fundamental phenomenon to the present application of proppant flows through intersecting fractures, the 100 mesh \ac{PSD} at $\bar{\phi}=0.5$ is firstly flowed through a straight, periodic channel (similar to that used in \citet{DiVaira2022} and the validation test in $\S$\ref{Section: Numerical model, Subsection: Validation} of the present work), and the particles are allowed to migrate and segregate until a steady state is reached. The \ac{PSD} of the particles in the near-wall region ($0<x_1<0.25w_0$) is then measured, which is plotted in Figure~\ref{fig_distributions_high-phi} against the original \ac{PSD} (i.e., over the entire channel). Therefore, in the results in $\S$\ref{Section: Realistic simulations, Subsection: Realistic results}, the entire simulation procedure described in $\S$\ref{Section: Realistic simulations, Subsection: Realistic test cell} and $\S$\ref{Section: Realistic simulations, Subsection: Simulation parameters} is repeated for this modified \ac{PSD}, at a higher \ac{SVF} of $\bar{\phi} = 0.4$ (which is the \ac{SVF} in the near-wall region when $\bar{\phi} = 0.5$ in the entire channel).
	
	\subsection{Results and analysis of realistic simulations}
	\label{Section: Realistic simulations, Subsection: Realistic results}
	
\begin{figure}
		\centering
		\subfloat[]{
			\includegraphics[width=0.55\linewidth]{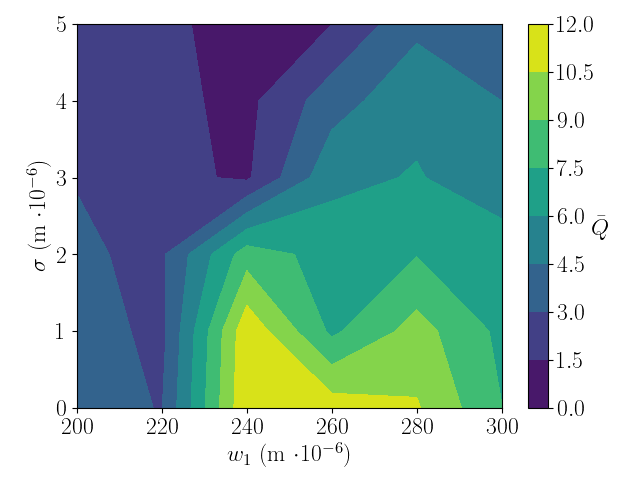}
			\label{fig_w_sigma_vel_base}
		}
		
		\vspace{-0.5em}
		\subfloat[]{
			\includegraphics[width=0.55\linewidth]{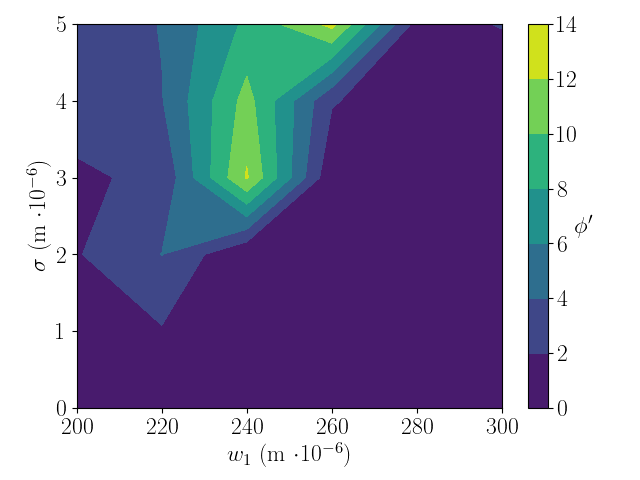}
			\label{fig_w_sigma_phi_base}	
		}
		
		\vspace{-0.5em}
		\subfloat[]{
			\includegraphics[width=0.55\linewidth]{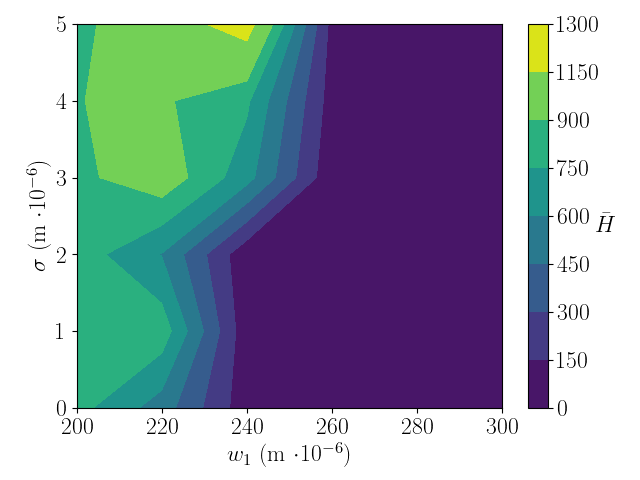}
			\label{fig_w_sigma_mound_base}	
		}
		\caption{Transport metrics of 100 mesh proppant at $\bar{\phi}=0.1$ flowing through rough cleats which intersect a hydraulic fracture, presented as contour maps of reference cleat width, $w_1$, versus cleat roughness, $\sigma$. The plotted metrics are the (a) non-dimensional flow rate, $\bar{Q}$, (b) retained \ac{SVF} in the cleat, $\phi'$, and (c) non-dimensional mound height, $\bar{H}$.}
		\label{fig_w_sigma_base}
	\end{figure}

	Figure~\ref{fig_w_sigma_base} presents the key metrics --- non-dimensional flow rate, $\bar{Q}$, retained \ac{SVF} in the cleat, $\phi'$, and non-dimensional mound height, $\bar{H}$ --- for 100 mesh proppant flowing through a rough cleat that intersects a hydraulic fracture. The results are plotted as contour maps in terms of two cleat parameters: the cleat width, $w_1$, and the cleat roughness, $\sigma$.
	
	Firstly, referring to Figure~\ref{fig_w_sigma_vel_base}, the flow rate is maximum for low cleat roughness ($\sigma \lesssim \SI{1}{\micro\metre}$) and cleat widths of $w_1 \gtrsim \SI{240}{\micro\metre}$, corresponding to the bottom-right corner of the figure. At these conditions, proppant flows directly through the cleat; Figure~\ref{fig_w_sigma_phi_base} shows that no proppant is retained in the cleat, and Figure~\ref{fig_w_sigma_mound_base} shows that no mounding occurs at the cleat entrance. At low roughness there is then a steep decrease in flow rate from $w_1=\SI{240}{\micro\metre}$ to \SI{220}{\micro\metre} (bottom-left corner of Figure~\ref{fig_w_sigma_vel_base}). Referring to Figure~\ref{fig_w_sigma_mound_base}, this directly corresponds to an increase in mounding, which occurs when occlusions are formed at the cleat entrance.
	
	Figure~\ref{fig_w_sigma_phi_base} demonstrates that proppant is increasingly retained in the cleat as the cleat roughness increases. Maximum retention occurs at $w_1 = \SI{240}{\micro\metre}$, which diminishes as the cleat becomes narrower or wider. Referring to Figure~\ref{fig_w_sigma_vel_base}, this maximum in retained proppant directly corresponds to a minimum in leak-off flow rate, which is commensurate with the STIM-LAB conclusions in $\S$\ref{Section: STIM-LAB experiment} and the simplified simulations in $\S$\ref{Section: Simplified simulations}. Mounding, on the other hand, primarily increases with decreasing $w_1$. The slight increase in mounding with $\sigma$ suggest that occlusions are only slightly more likely to form when the cleat roughness is higher, but that the overriding determinant of occlusion formation is the cleat aperture. Additionally, even when no proppant retention or mounding occurs ($w_1 > \SI{260}{\micro\metre}$), the leak-off rate uniformly decreases with increasing $\sigma$, indicating an increase in resistance to the flow due to the surface roughness of the cleat.
	
	In general, high-fidelity maps like Figure~\ref{fig_w_sigma_vel_base} can be directly incorporated into \ac{HF} simulators to give an improved prediction of fluid leak-off, if cleat widths and roughness are approximately known. While Figure~\ref{fig_w_sigma_vel_base} accounts only for constant leak-off rates, neglecting decreases due to filter-caking or reservoir pressure dependence, it provides a datum on which existing leak-off predictive models (such as the Carter leak-off equation) can be modified to provide significantly improved approximations. Furthermore, while the map here is for 100 mesh proppant only, it could be regenerated for any size distribution, as will be demonstrated next, or for any ratio of proppant sizes to cleat sizes.
	
\begin{figure}
		\centering
		\subfloat[]{
			\includegraphics{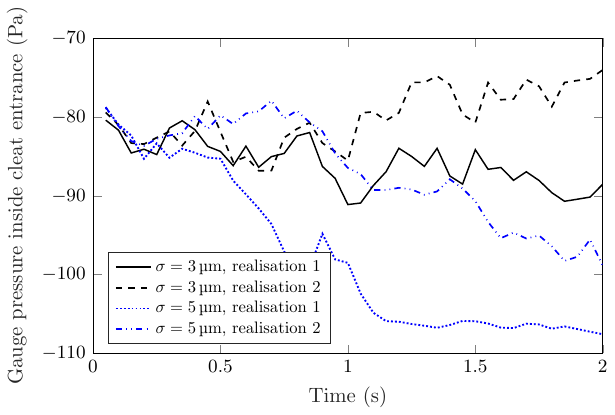}
			\label{fig_params_vs_time_pressure}
		}
		
		\vspace{-0.5em}
		\subfloat[]{
			\includegraphics{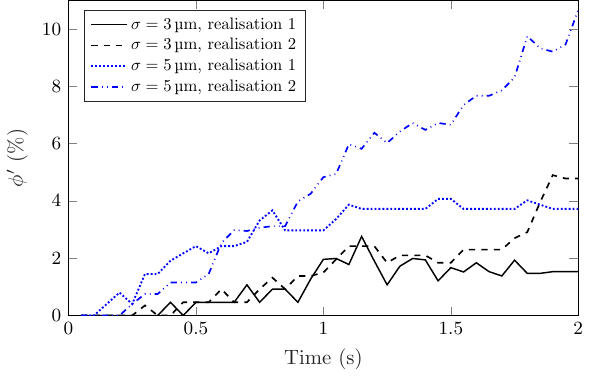}
			\label{fig_params_vs_time_phi}	
		}
		
		\vspace{-0.5em}
		\subfloat[]{
			\includegraphics{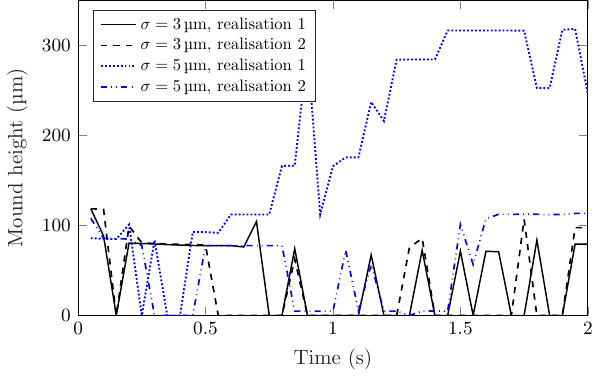}
			\label{fig_params_vs_time_mound}	
		}
		\caption{Interrogation of (a) the pressure just inside the cleat entrance, alongside (b) the \ac{SVF} retained in the cleat, $\phi'$, and (c) the non-dimensional mound height, $\bar{H}$, for four different simulations (two realisations each for $\sigma = \SI{3}{\micro\metre}$ and \SI{5}{\micro\metre}, all at $w_1 = \SI{240}{\micro\metre}$).}
		\label{fig_params_vs_time}
	\end{figure}

\begin{figure}
		\centering
		\subfloat[]{
			\includegraphics[width=0.2\linewidth]{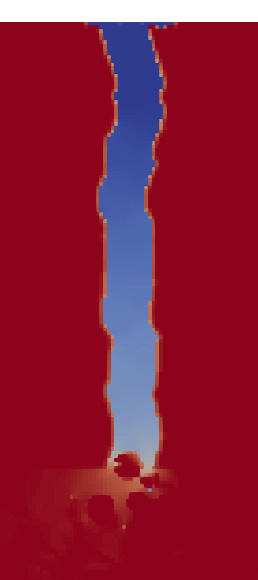}
			\label{fig_pressure_a}
		}
		\hspace{2em}
		\subfloat[]{
			\includegraphics[width=0.2\linewidth]{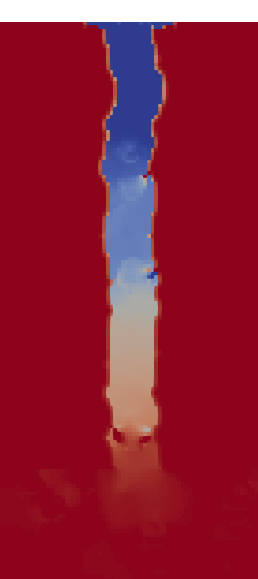}
			\label{fig_pressure_b}	
		}
		
		\subfloat{
			\includegraphics[width=0.35\textwidth, height=5mm]{fig_colourmap_redblue.png}
		}
		
		{\small \hspace{-1em} -138 \hspace{1.5em} Gauge pressure (\SI{}{\pascal}) \hspace{1.5em} 0}
		\caption{Pressure distributions through the cleat and cleat entrance for (a) realisation 1 and (b) realisation 2 of $\sigma = \SI{5}{\micro\metre}$ and $w_1 = \SI{240}{\micro\metre}$ (from Figure~\ref{fig_params_vs_time}). Distributions are captured at $t=\SI{2}{\second}$, with the slice in the $x_1-x_2$ plane at $h/2$.}
		\label{fig_pressure}
	\end{figure}

	To gain further physical insight into the results, the fluid pressure is interrogated at a point just inside the cleat entrance for a subset of simulations (two realisations each for $\sigma = \SI{3}{\micro\metre}$ and \SI{5}{\micro\metre} at $w_1 = \SI{240}{\micro\metre}$), and is plotted in Figure~\ref{fig_params_vs_time} alongside the retained \ac{SVF} and mound height. For realisation 1 at $\sigma = \SI{5}{\micro\metre}$, an occlusion is immediately formed at the cleat entrance, leading to the formation of a large mound (Figure~\ref{fig_params_vs_time_mound}). This causes a significant pressure drop through the mound, and therefore a low pressure just inside the cleat entrance (Figure~\ref{fig_params_vs_time_pressure}), as described in the analysis from $\S$\ref{Section: Simplified simulations, Subsection: Results}. Conversely, for realisation 2 at $\sigma = \SI{5}{\micro\metre}$, a smaller mound forms but a high \ac{SVF} of particles also eventually become retained in the cleat (Figure~\ref{fig_params_vs_time_phi}). This means that more of the pressure drop occurs through the cleat compared to realisation 1, and that the pressure at the cleat entrance is comparatively higher. These pressure drops are also visualised in the pressure distribution plots of Figure~\ref{fig_pressure}. For both of the realisations at $\sigma = \SI{3}{\micro\metre}$, on the other hand, no proppant builds up at the cleat entrances, meaning that all of the pressure drop occurs through the cleat only, resulting in the pressure staying high at the cleat entrance. The pressure is higher for realisation 2 compared to realisation 1 seeing as more proppant has been retained in the cleat, resulting in an even greater pressure drop from the cleat exit to the cleat entrance.
	
\begin{figure}
		\centering
		\subfloat[]{
			\includegraphics[width=0.55\linewidth]{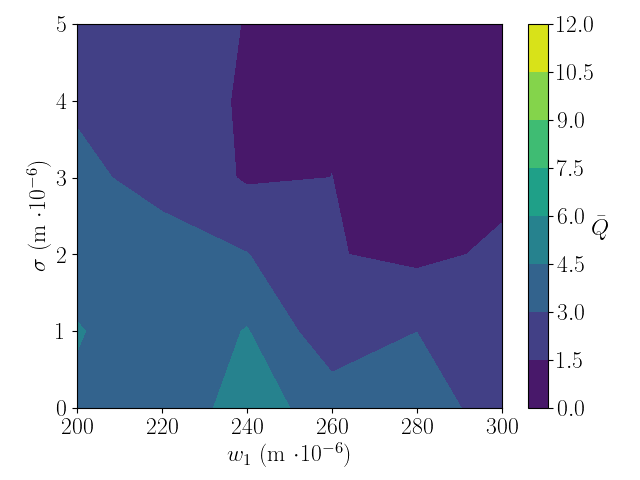}
			\label{fig_w_sigma_vel_high-phi}
		}
		
		\vspace{-0.5em}
		\subfloat[]{
			\includegraphics[width=0.55\linewidth]{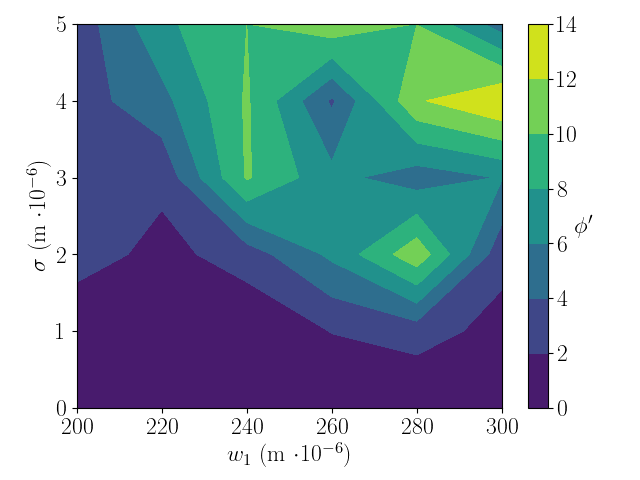}
			\label{fig_w_sigma_phi_high-phi}	
		}
		
		\vspace{-0.5em}
		\subfloat[]{
			\includegraphics[width=0.55\linewidth]{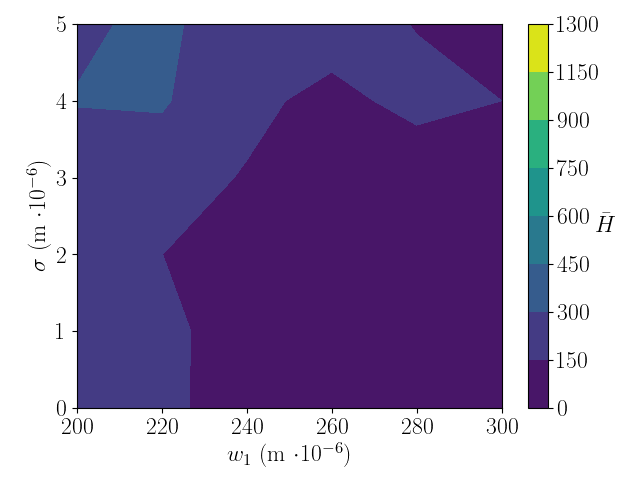}
			\label{fig_w_sigma_mound_high-phi}	
		}
		\caption{Transport metrics of 100 mesh with size distribution modified due to migration effects, at $\bar{\phi}=0.4$, presented as contour maps of reference cleat width, $w_1$, versus cleat roughness, $\sigma$. The plotted metrics are the (a) non-dimensional flow rate, $\bar{Q}$, (b) retained \ac{SVF} in the cleat, $\phi'$, and (c) non-dimensional mound height, $\bar{H}$.}
		\label{fig_w_sigma_high-phi}
	\end{figure}

	Next, Figure~\ref{fig_w_sigma_high-phi} plots the results for the modified 100 mesh \ac{PSD}, which is representative of the near-wall region when particles have segregated at $\bar{\phi}=0.5$ due to shear-induced migration. Due to this segregation, the migrated \ac{PSD} contains a higher proportion of larger particles compared to the original \ac{PSD} (Figure~\ref{fig_distributions_high-phi}). Firstly, Figure~\ref{fig_w_sigma_vel_high-phi} demonstrates that leak-off is lower over all $\sigma$ and $w_1$ compared to the original distribution. One reason for this is that proppant has been retained in the cleat over a much wider range of $\sigma$ and $w_1$ (Figure~\ref{fig_w_sigma_phi_high-phi}), due to the higher proportion of larger particles. Interestingly, the magnitude of the maximum $\phi'$ is the same, even though the injected $\bar{\phi}$ is a factor of four higher. However, even when no proppant has been retained in the cleat, the leak-off rate is lower than the original distribution (refer to bottom-left corners of Figures~\ref{fig_w_sigma_vel_high-phi}~and~\ref{fig_w_sigma_phi_high-phi}). This suggests that a higher $\bar{\phi}$ results in lower leak-off. Figure~\ref{fig_w_sigma_mound_high-phi} shows that mounding is also significantly lower compared to the original case. This most likely results from the higher $\bar{\phi}$ in the primary channel disturbing mounds to a greater degree. Overall, re-running the simulations with this modified \ac{PSD} represents just one of the many modifications which can be made to the present methodology. Depending on the specific reservoir and suspension characteristics, tailored data could be obtained for different proppant size ranges, cleat sizes, fluid characteristics, and even additional physics, such as electrostatics.
	
	\section{Conclusions}
	
	This work simulates the transport of proppant through coal cleats at the point which they intersect a hydraulic fracture, for the first time accounting for the realities of proppant retention and mounding due to variability in cleat sizes. Using two sets of simulations, one simplified and the other more realistic, novel insights and data are generated which can immediately inform the design of future proppant treatments.
	
	In the simplified simulations, it is demonstrated that fluid leak-off is minimised when proppant invades and is retained in the cleats. This occurs when the cleat aperture is sufficiently wide to admit proppant. As such, for the cleat widths tests, significant mounding and leak-off occurs for 100 mesh and 40/70 mesh. However, when proppant of a wide size distribution is used, such as 100/635 mesh, leak-off and mounding are significantly reduced. These findings are commensurate with the benchmark experiment of \citet{Stimlab1995}. Here, for the first time, it is shown that shear-thinning fluids result in significantly less mounding than Newtonian fluids for similar leak-off control. All of these findings are given physical and mathematical explanations by relating the flow through the proppant mounds and propped cleats to flow through a porous medium.
	
	Realistic simulations are then conducted which model the cleat with physical roughness. Firstly, it is demonstrated that fluid leak-off is greatest for low cleat roughness and above some critical cleat width. At these conditions, proppant flows directly through the cleat. As the cleat width decreases below the critical width, leak-off quickly decreases. This occurs as occlusions increasingly form at the cleat entrance, which also leads to increasing mounding. However, occlusions are only slightly more likely at high roughness compared to low roughness. In general, the minimum leak-off occurs when proppant retention in the cleat is maximised. This occurs approximately at the critical cleat width, but above some critical cleat roughness. When no retention occurs (above the critical width), but for high roughness, the leak-off rate is smaller compared to low roughness, due to increased resistance to flow. Finally, the critical width shifts depending on the shape of the particle size distribution, as does the range of width and roughness over which retention occurs. For higher injected solid volume fractions of proppant, leak-off is reduced.
	
	Overall, the maps which are generated in terms of the cleat width and roughness could be used to modify existing leak-off predictive models, as well as improve predictions of propped reservoir volumes, for incorporation into hydraulic fracturing simulators. The results can be tailored for the desired treatment conditions, such as different proppant sizes and distributions, cleat sizes, fluid characteristics, and even additional physics, such as electrostatics.
	
	\section*{Acknowledgements}
	
	The authors gratefully acknowledge the support of this work, including the Advance Queensland Industry Research Fellowship scheme (AQIRF1372018), National Energy Resources Australia (NERA), the University of Queensland (UQ) via an Australian Government Research Training Program Scholarship, and the proponents of the UQ Gas and Energy Transition Research Centre (Australia Pacific LNG, Santos, and Arrow Energy). This work was undertaken with the assistance of resources and services from the National Computational Infrastructure (NCI) and the Pawsey Supercomputing Centre, each with funding from the Australian Government, with the latter also funded by the Government of Western Australia.

	\clearpage
	
	\bibliographystyle{elsarticle-harv} 
	\bibliography{bibliography.bib}

\end{document}